\documentclass[twocolumn, aps, prc, superscriptaddress, showpacs,
floatfix]{revtex4}

\usepackage{multirow}
\usepackage{epsfig}
\usepackage{amsmath}

\usepackage{subfigure}
\usepackage{latexsym}
\usepackage{feynmp}

\usepackage[normalem]{ulem}  
\usepackage[dvips]{color} 

\renewcommand\sout{\bgroup \color{red} \ULdepth=-.5ex \ULset}

\begin{document}

\title{Hadronic effects on the X(3872) meson abundance in heavy
ion collisions}

\author{Sungtae Cho}
\author{Su Houng Lee}
\affiliation{Institute of Physics and
Applied Physics, Yonsei University, Seoul 120-749, Korea}

\begin{abstract}

We study the hadronic effects on the $X(3872)$ meson abundance in
heavy ion collisions. We evaluate the absorption cross sections of
the $X(3872)$ meson by pions and rho mesons in the hadronic stage
of heavy ion collisions, and further investigate the variation in
the $X(3872)$ meson abundance during the expansion of the hadronic
matter for its two possible quantum number states; $J^P=1^+$ and
$2^-$. We show that the absorption cross sections and the time
evolution of the $X(3872)$ meson abundance are strongly dependent
on the structure and quantum number of the $X(3872)$ meson. We
thus suggest that studying the abundance of the $X(3872)$ meson in
relativistic heavy ion collisions provides a chance to infer its
quantum number as well as its structure.

\end{abstract}


\maketitle

\section{Introduction}


Relativistic heavy ion collision experiments have made it possible
to study a system of quantum chromodynamic (QCD) matter at very
high temperature and density in the laboratory
\cite{Arsene:2004fa, Back:2004je,Adams:2005dq,Adcox:2004mh,
Gyulassy:2004zy}. The research on the system of deconfined quarks
and gluons, so-called quark-gluon plasma (QGP), has also enabled
us to understand the possible phase transition between the hot and
dense matter and QGP \cite{Gupta:2011wh}. Moreover, due to the
enormous energies produced in heavy ion collisions, particles that
are otherwise hard to find in nature could be produced during the
quark-hadron phase transition.

Recently STAR Collaboration reported the observation of an
antimatter helium-4 nucleus as well as an antimatter hypernucleus
produced at RHIC \cite{Abelev:2010rv, Agakishiev:2011ib}, and also
tried to measure the signal of an exotic H dibaryon
\cite{Shah:2011np}. As attempts to understand the production of
particles of these kinds, there have been many studies focusing on
their production yields based on both the statistical model and
the coalescence model \cite{Andronic:2010qu, Cho:2010db,
Cho:2011ew, Cleymans:2011pe, Xue:2012gx, Steinheimer:2012tb}.
Moreover, as one of the possible methods to understand the
structure of the exotic hadrons, a new approach of studying exotic
hadrons in relativistic heavy ion collision experiments has been
proposed \cite{Cho:2010db,Cho:2011ew}. There, the relation between
the production yields of exotic hadrons and the structure at the
moment of their formation has been sought out by considering the
production of all proposed possible structures using the
coalescence model, and it was found that the production yields of
exotic hadron candidates strongly reflect their structures.

The abundance of hadrons evaluated at the chemical freeze-out
temperature, however, may change due to the dissociation or the
absorption by mostly light mesons such as the pion and the $\rho$
meson in the hadronic medium. The effects from the hadronic
interactions on the production of heavy quark mesons have been
discussed in many literatures. In order to estimate the
possibilities of $J/\psi$ suppression in the hadronic matter, one
meson exchange model with the effective Lagrangian has been
introduced to evaluate the absorption cross sections with light
hadrons \cite{Matinyan:1998cb, Haglin:1999xs, Lin:1999ad,
Oh:2000qr}. A similar approach has been applied to investigate the
time evolution of $D_{sJ}(2317)$ meson abundance in the hadronic
matter \cite{Chen:2007zp}. In this work we investigate the
hadronic medium effects on the production yield of one of exotic
mesons, the $X(3872)$ meson.

The $X(3872)$ meson was first discovered by Belle Collaboration
\cite{Choi:2003ue} from the measurement of $B^+\to
J/\Psi\pi^+\pi^-K^+$ decay, and later confirmed by CDF
\cite{Acosta:2003zx}, D0 \cite{Abazov:2004kp}, and BABAR
\cite{Aubert:2004ns} collaborations. The additional decay modes of
$X(3872)$ mesons to $D^0\bar{D}^0\pi^0$
\cite{Gokhroo:2006bt,Aubert:2007rva,Adachi:2008sua},
$J/\Psi\omega$ \cite{delAmoSanchez:2010jr}, $J/\Psi\gamma$
\cite{Aubert:2006aj}, and $\Psi(2s)\gamma$ \cite{Aubert:2008ae}
have also been observed. The positive charge parity of the
$X(3872)$ meson has been established by the observation of the
$X(3872)$ meson decaying to $J/\Psi\gamma$ \cite{Aubert:2006aj}
and $\Psi(2s)\gamma$ \cite{Aubert:2008ae}, and the current world
average mass of the $X(3872)$ meson in PDG \cite{Beringer:1900zz}
is 3871.68 $\pm$ 0.17 MeV. However, it is still not clear what the
exact structure and quantum number of the $X(3872)$ meson is.
Suggested hypotheses for the structure of the $X(3872)$ meson
include a pure charmonium state, a $\bar{D}^0D^{*0}$ hadronic
molecule, a tetra-quark state, and a charmoniun-gluon hybrid state
\cite{Nielsen:2009uh}. From the analysis of the angular
distribution of the $X(3872)$ meson decaying to $J/\Psi\pi^+\pi^-$
\cite{Abulencia:2006ma}, we now understand that the possible
quantum number $J^{P}$ should be either $1^+$ or $2^-$.

There are various experimental results supporting each spin
possibility of the $X(3872)$ meson. The study of the $X(3872)$
meson decaying to $\bar{D}^0D^{*0}$ disfavors the $2^-$ quantum
number because of the angular momentum barrier in its
near-threshold decay \cite{Gokhroo:2006bt,Aubert:2007rva}. On the
other hand, the analysis of the $X(3872)$ meson decaying to
$J/\Psi\omega$ favors a $p$ wave in the final state, $2^-$
\cite{delAmoSanchez:2010jr}. We expect that the two different spin
possibilities of the $X(3872)$ meson will also lead to different
experimental results in heavy ion collision experiment.

After the $X(3872)$ meson is produced at the chemical freeze-out,
it interacts with other hadrons during the expansion of the
hadronic matter. As a result, the $X(3872)$ meson can be absorbed
by the comoving light mesons or additionally produced from
interactions between charmed mesons such as $D$ and $\bar{D}^*$.
Evaluating the $X(3872)$ meson cross sections by light hadrons
therefore should be useful in estimating the hadronic effects on
the $X(3872)$ meson abundance in heavy ion collisions. However,
the $X(3872)$ meson would interact with light hadrons differently
depending on the spin of the $X(3872)$ meson. In order for the
spin-2 $X(3872)$ meson to interact with light mesons, there should
be a charmed meson having the relative momentum to satisfy the
angular momentum conservation. Also the spin-2 $X(3872)$ meson
should carry a symmetric traceless spin polarization tensor
whereas the spin-1 $X(3872)$ meson carries a polarization vector.
Therefore, we expect to obtain two different results when we
evaluate the cross sections of the $X(3872)$ meson for the two
different spin states. By comparing these results with the
experimental observation in heavy ion collisions, we may obtain a
hint for the quantum number of the $X(3872)$ meson.

In this study we restrict our consideration of the $X(3872)$ meson
structure to the spin-1 tetra-quark state and the spin-2 pure
charmonium state. We briefly discuss the $\bar{D}^0D^{*0}$
hadronic molecule for the spin-1 $X(3872)$ meson possibly produced
during the hadronic stage and at the kinematical freeze-out point.
All the discussion will be focused on the central heavy ion
collisions at Relativistic Heavy Ion Collider (RHIC) at Brookhaven
National Laboratory; using a model developed to describe the
dynamics of the cental Au+Au collisions at $\sqrt{s_{NN}}$ = 200
GeV. Hereafter, we use simplified notations for the $X(3872)$
meson; $X_1$ for a $1^+$ state and $X_2$ for a $2^-$ state.

This paper is organized as follows. In Sec. II, we briefly discuss
the production of the $X(3872)$ meson at the chemical freeze-out
in both the statistical and coalescence model. Then we consider
the interaction of the $X(3872)$ meson with light mesons such as
pions and $\rho$ mesons, and evaluate the cross sections of the
$X(3872)$ meson in the hadronic medium in Sec. III. In Sec. IV we
investigate the time evolution of the $X(3872)$ meson abundance by
solving the kinetic equation based on the phenomenological model.
Section V is devoted to conclusions. In Appendix A, we discuss the
dependence of the strong-coupling constants on the $X(3872)$ meson
mass. We briefly address the hadronic effects on the $X(3872)$
meson by baryons in Appendix B.

\section{$X(3872)$ meson production from the quark-gluon plasma}

We evaluate the production yields of the $X(3872)$ meson in heavy
ion collisions using both the statistical and the coalescence
model. The statistical model, which assumes that hadrons are in
thermal and chemical equilibrium when they are produced at
chemical freeze-out in heavy ion collisions, has been very
successful in describing the production yields of hadrons
\cite{BraunMunzinger:1994xr, BraunMunzinger:1995bp,
BraunMunzinger:1999qy, BraunMunzinger:2001ip}. We apply the same
parameters evaluated in Ref. \cite{Cho:2011ew} to obtain the
thermal yields.

\allowdisplaybreaks{
\begin{eqnarray} && N_X^\mathrm{stat} = V_H
\frac{g_X}{2 \pi^2} \int_0^\infty \frac{p^2
dp}{\gamma_C^{-1}e^{E_X/T_H} \pm 1}
\nonumber \\
&& \quad\quad~\approx
\frac{\gamma_Cg_XV_H}{2\pi^2}m_X^2T_HK_2(m_X/T_H), \label{Stat}
\end{eqnarray} }
where the Maxwell-Boltzmann approximation has been made in the
second line. We consider the $X(3872)$ meson produced at the
hadronization temperature $T_H$ = 175 MeV when the volume of the
quark-gluon plasma, $V_H$ is 1908 fm$^3$ \cite{Cho:2010db,
Cho:2011ew}. We assume that the total number of charm quark
produced from the initial hard collisions at RHIC is 3, which
leads to the charm quark fugacity factor $\gamma_C$ = 6.4 by the
requirement that the charm quark is conserved among charmed
hadrons such as $D$, $D^*$, $D_s$ mesons, and $\Lambda_c$. The
difference in the yields between the spin-1 $X(3872)$ meson and
the spin-2 $X(3872)$ meson originates only from the spin
degeneracy $g_X$ in the statistical model.

In the coalescence model, which successfully explains both the
enhancement of the baryon to meson ratio in the intermediate
transverse momentum region \cite{Greco:2003xt, Greco:2003mm,
Fries:2003vb, Fries:2003kq} and the quark number scaling of the
elliptic flows \cite{Molnar:2003ff}, we consider the yields of the
$X(3872)$ meson produced from both the four quark configuration
for the spin-1 state and the two quark configuration for the
spin-2 state. We assume that the quark coalescence occurs in the
volume 1000 fm$^3$ and the mass of light constituent quarks is 300
MeV, while that of a charm constituent quark is 1500 MeV. We also
assume that the available light quark number at hadronization
temperature is 245. We adopt the oscillator frequency of the
Wigner function for charmed hadrons $\omega_c$ = 385 MeV obtained
by requiring that the coalescence model reproduces well the yield
of $\Lambda_c(2286)$ in the statistical model. For details, refer
to Ref. \cite{Cho:2011ew}. We summarize the production yields of
the $X(3872)$ meson in Table \ref{yields}.

\begin{table}[!h]
\caption{The $X(3872)$ meson yields in central Au+Au collisions at
$\sqrt{s_{NN}}=200$ GeV at RHIC in both the statistical and
coalescence model. The yields for the spin-2 state differ from
those in Table V. of Ref. \cite{Cho:2011ew} by the spin degeneracy
factor 5/3.} \label{yields}
\begin{center}
\begin{tabular}{cccc}
\hline \hline
$ $X(3872)$ $ & Coal.(2q) & Coal.(4q) & Stat. \\
\hline
spin-1 & & $ \quad 4.0\times 10^{-5} $ & $ \quad 2.9\times 10^{-4} $ \\
spin-2 & $ \quad 1.7\times 10^{-4} $ & & $ \quad 4.8\times 10^{-4} $ \\
\hline \hline
\end{tabular}
\end{center}
\end{table}
In Table \ref{yields}, the smaller yields in the coalescence model
compared to those in the statistical model reflect the suppression
effects in the quark coalescence process. The suppression
mechanism is, however, different for different spin states of the
$X(3872)$ meson. The coalescence of additional quarks to construct
the tetraquark state makes the yield suppressed for the spin-1
$X(3872)$ meson whereas the construction of a $d$-wave coalescence
factor leads to the smaller yield for the spin-2 $X(3872)$ meson.

\begin{widetext} 

\begin{figure}[t]
\begin{center}
\subfigure[]{
\includegraphics[width=0.17\textwidth]{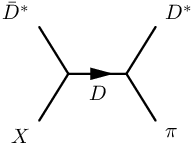}}
\quad \subfigure[]{
\includegraphics[width=0.17\textwidth]{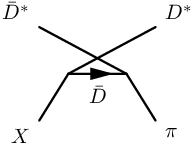}}
\quad \subfigure[]{
\includegraphics[width=0.16\textwidth]{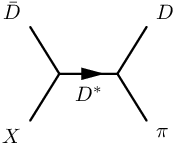}}
\quad \subfigure[]{
\includegraphics[width=0.16\textwidth]{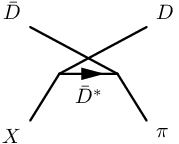}}
\quad \subfigure[]{
\includegraphics[width=0.17\textwidth]{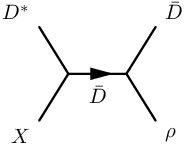}}
\subfigure[]{
\includegraphics[width=0.17\textwidth]{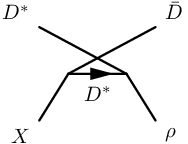}}
\quad \subfigure[]{
\includegraphics[width=0.17\textwidth]{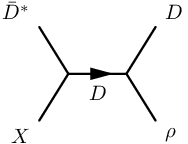}}
\quad \subfigure[]{
\includegraphics[width=0.17\textwidth]{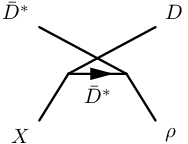}}
\quad \subfigure[]{
\includegraphics[width=0.16\textwidth]{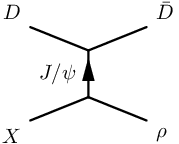}}
\quad \subfigure[]{
\includegraphics[width=0.17\textwidth]{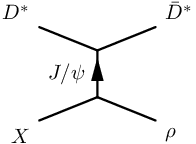}}
\end{center}
\caption{Born diagrams for $X(3872)$ absorption by pions and
$\rho$ mesons: $X\pi\to \bar{D}^*D^*$, (a) and (b); $X\pi\to
\bar{D}D$, (c) and (d); $X\rho\to D^*\bar{D}$, (e) and (f);
$X\rho\to \bar{D}^*D$, (g) and (h); $X\rho\to D\bar{D}$, (i); and
$X\rho\to D^*\bar{D}^*$, (j).} \label{Diagrams}
\end{figure}
\end{widetext}

\section{Hadronic effects on the $X(3872)$ meson}

The $X(3872)$ meson produced at the chemical freeze-out interacts
with other hadrons during the expansion of the hadronic matter. As
a result the $X(3872)$ meson can be absorbed by the comoving light
mesons or produced from the interaction between charmed mesons
such as $D$ and $\bar{D}^*$. We consider here the $X(3872)$ meson
interacting with light mesons such as pions and $\rho$ mesons,

\begin{eqnarray}
& & X\pi\rightarrow \bar{D}^*D^*, \quad~ X\pi \rightarrow
\bar{D}D,\quad~X\rho \rightarrow  \bar{D}^*D, \nonumber \\
& & X\rho \rightarrow D^*\bar{D}, \qquad X\rho \rightarrow
D\bar{D}, \quad~ X\rho \rightarrow D^*\bar{D}^*. \label{processes}
\end{eqnarray}
The diagrams representing each process in Eq. (\ref{processes})
are shown in Fig. \ref{Diagrams}. To evaluate the cross sections
for these diagrams in Fig. \ref{Diagrams}, we consider the
following interaction Lagrangians

\begin{eqnarray} {\cal L}_{\pi DD^*}&=&ig_{\pi
DD^*}D^{*\mu}\vec\tau\cdot(\bar D
\partial_\mu\vec\pi-\partial_\mu\bar D\vec\pi) + {\rm H.c.},
\nonumber \\
{\cal L}_{\rho DD}&=&ig_{\rho DD}(D\vec\tau\partial_\mu\bar D
-\partial_\mu D\vec\tau\bar D)\cdot\vec\rho^\mu,
\nonumber \\
{\cal L}_{\rho D^*D^*}&=&ig_{\rho D^*D^*}~[ (\partial_\mu D^{*\nu}
\vec\tau\bar{D^*_\nu}-D^{*\nu}\vec\tau\partial_\mu\bar{D^*_\nu})
\cdot\vec\rho^\mu \nonumber \\
&+& (D^{*\nu}\vec\tau\cdot\partial_\mu\vec\rho_\nu-\partial_\mu
D^{*\nu}\vec\tau\cdot\vec\rho_\nu)\bar {D^{*\mu}} \nonumber \\
&+& D^{*\mu}(\vec\tau\cdot\vec\rho^\nu\partial_\mu\bar{D^*_\nu}
-\vec\tau\cdot\partial_\mu\vec\rho^\nu \bar {D^*_\nu}) ], \nonumber \\
{\cal L}_{\psi DD}&=&ig_{\psi DD}\psi^\mu(D\partial_\mu\bar D-
\partial_\mu D\bar D), \nonumber \\
{\cal L}_{\psi D^*D^*}&=& ig_{\psi D^*D^*}[\psi^\mu(\partial_\mu
D^{*\nu}\bar{D^*_\nu}-D^{*\nu}\partial_\mu\bar {D^*_\nu})
\nonumber \\
&+& (\partial_\mu\psi^\nu D^*_\nu-\psi^\nu\partial_\mu D^*_\nu)
\bar{D^{*\mu}} \nonumber \\
&+& D^{*\mu}(\psi^\nu\partial_\mu\bar{D^*_\nu}-\partial_\mu
\psi^\nu\bar{D^*_\nu}) ] . \label{fLagrangians}
\end{eqnarray}
In Eq. (\ref{fLagrangians}), $\vec \tau$ are the Pauli matrices,
and $\vec \pi$ and $\vec \rho$ denote the pion and rho meson
isospin triplets, respectively, while $D\equiv (D^0,D^+)$ and
$D^*\equiv (D^{*0},D^{*+})$ denote the pseudoscalar and vector
charm meson doublets, respectively. Here the shorthand notation
$\psi$ has been used for the $J/\psi$ meson. These interaction
Lagrangians have been obtained from the free Lagrangians for
pseudoscalar and vector mesons by introducing the minimal
substitution \cite{Lin:1999ad}. On the other hand, the interaction
Lagrangians for the $X(3872)$ meson have been built to produce
strong transition matrix elements for the $X(3872)$ meson decays;
X$\to J/\psi\rho$ and X$\to D^0\bar{D}^{*0}$ \cite{Brazzi:2011fq},

\begin{eqnarray}
{\cal L}_{X_1D^*D}&=&g_{X_1D*D}X_1^\mu \bar{D}_\mu^*D, \nonumber \\
{\cal L}_{X_1\psi\rho}&=&ig_{X_1\psi\rho}\epsilon^{\mu\nu\rho
\sigma}\psi_\nu\rho_\rho\partial_\sigma X_{1\mu},  \nonumber \\
{\cal L}_{X_2D^*D}&=&-ig_{X_2D^*D}X_2^{\mu\nu}\bar{D}_\mu^*
\partial_\nu D, \nonumber \\
{\cal L}_{X_2\psi\rho}&=&-g_{X_2\psi\rho}\epsilon^{\mu\nu\rho
\sigma}X_{2\mu\alpha}(\partial_\nu\psi^{\alpha}\partial_\rho
\rho_\sigma-\partial_\nu\rho^{\alpha}\partial_\rho \psi_\sigma)
\nonumber \\
&+& g_{X_2\psi\rho}'\epsilon^{\mu\nu\rho\sigma}\partial_\nu
X_{2\mu\alpha}(\partial^\alpha\psi_{\rho}\rho_\sigma-\psi_\rho
\partial^\alpha \rho_\sigma). \label{iLagrangians}
\end{eqnarray}

As we see in Eq. (\ref{iLagrangians}), we need an additional
derivative for the interaction Lagrangians of the spin-2 $X(3872)$
meson compared to those of the spin-1 $X(3872)$ meson to satisfy
the angular momentum conservation. The structure of these
Lagrangians is expected to make the energy dependence on the cross
sections different in the hadronic medium. This is also the factor
prohibiting the spin-2 $X(3872)$ meson from decaying to the vector
meson $D^*$ and the pseudoscalar meson $D$ near the threshold
energy. We need an anti-symmetric tensor
$\epsilon^{\mu\nu\rho\sigma}$ to describe the isospin violating
interaction between one axial vector meson and two vector mesons
for the spin-1 $X(3872)$ meson. This term becomes more complicated
for the spin-2 $X(3872)$ meson since the anti-symmetric tensor
$\epsilon^{\mu\nu\rho\sigma}$ should not be fully contracted with
the symmetric polarization tensor $\pi_{\mu\nu}$.

Based on the above effective Lagrangians, we consider the
reactions for the $X(3872)$ meson absorption by pions and $\rho$
mesons shown in Fig. \ref{Diagrams}. Among those diagrams in Fig.
\ref{Diagrams} the process $X\rho \rightarrow \bar{D^*}D $ leads
to the same cross section as the process $ X\rho \rightarrow
D^*\bar{D} $. The amplitudes for all processes, without isospin
factors and before summing and averaging over external spins, are
represented by

{\allowdisplaybreaks
\begin{eqnarray} {\cal M}_{X\pi\rightarrow \bar{D}^*D^*} &\equiv&
{\cal M}^{(a)}_{X}+{\cal M}^{(b)}_{X}, \nonumber \\
{\cal M}_{X\pi\rightarrow \bar{D}D} &\equiv& {\cal
M}^{(c)}_{X}+{\cal M}^{(d)}_{X}, \nonumber \\
{\cal M}_{X\rho\rightarrow D^*\bar{D}} &\equiv& {\cal
M}^{(e)}_{X}+{\cal M}^{(f)}_{X}, \nonumber \\
{\cal M}_{X\rho\rightarrow \bar{D}^*D} &\equiv& {\cal
M}^{(g)}_{X}+{\cal M}^{(h)}_{X}, \nonumber \\
{\cal M}_{X\rho\rightarrow D\bar{D}} &\equiv& {\cal
M}^{(i)}_{X}, \nonumber \\
{\cal M}_{X\rho\rightarrow D^*\bar{D}^*} &\equiv& {\cal
M}^{(j)}_{X}, \label{amplitudes}
\end{eqnarray} }
where the amplitudes for the first process $X\pi\rightarrow
\bar{D^*}D^*$ are

\begin{eqnarray} &&{\cal
M}^{(a)}_{X_1}=-g_{\pi D^*D}g_{X_1 D^*D}\epsilon^{\mu}_{1}
\epsilon^{\nu}_{2}\epsilon_{3\mu}^*\frac{1}{t-m_D^2}(2p_2-p_4)_\nu,
\nonumber \\
&&{\cal M}^{(b)}_{X_1}=-g_{\pi D^*D}g_{X_1 D^*D}\epsilon^{\mu}_{1}
\epsilon^{\nu}_{2}\epsilon_{3\mu}^*\frac{1}{u-m_D^2}(2p_2-p_3)_\nu
\nonumber \\ \label{Matx_X1pi1}
\end{eqnarray}
for the spin-1 $X(3872)$ meson and

\begin{eqnarray}
&&{\cal M}_{X_2}^{(a)}=g_{\pi D^*D}g_{X_2 D^*D}\pi^{\mu\alpha}_{1}
\epsilon^{\nu}_{2}\epsilon_{3\mu}^* \nonumber \\
&&\quad\quad~\times\frac{1}{t-m_D^2}(p_4-p_2)_\alpha(2p_2-p_4)_\nu,
\nonumber \\
&&{\cal M}_{X_2}^{(b)}=-g_{\pi D^*D}g_{X_2 D^*D}\pi^{\mu\alpha}_{1}
\epsilon^{\nu}_{2}\epsilon_{3\mu}^* \nonumber \\
&&\quad\quad~\times\frac{1}{u-m_D^2}(p_3-p_2)_\alpha(2p_2-p_3)_\nu
\label{Matx_X2pi1}
\end{eqnarray}
for the spin-2 $X(3872)$ meson. Similarly, the amplitudes for the
second process $X\pi \rightarrow \bar{D}D$ are

\begin{eqnarray}
&&{\cal M}_{X_1}^{(c)}=g_{\pi D^*D}g_{X_1 D^*D}\epsilon^{\mu}_{1}
\frac{1}{t-m_{D^*}^2}(p_2+p_4)^\nu \nonumber \\
&&\quad\quad~\times\Big(-g_{\mu\nu}+\frac{(p_1-p_3)_\mu(p_1-
p_3)_\nu}{m_{D^*}^2}\Big), \nonumber \\
&&{\cal M}_{X_1}^{(d)}=-g_{\pi D^*D}g_{X_1 D^*D}\epsilon^{\mu}_{1}
\frac{1}{u-m_{D^*}^2}(p_2+p_3)^\nu\nonumber \\
&&\quad\quad~\times\Big(-g_{\mu\nu}+\frac{(p_1-p_4)_\mu(p_1-
p_4)_\nu}{m_{D^*}^2}\Big) \label{Matx_X1pi2}
\end{eqnarray}
and

\begin{eqnarray}
&&{\cal M}_{X_2}^{(c)}=-g_{\pi D^*D}g_{X_2 D^*D}\pi^{\mu\alpha
}_{1}\frac{1}{t-m_{D^*}^2}(p_2+p_4)^\nu p_{3\alpha} \nonumber \\
&&\quad\quad~\times\Big(-g_{\mu\nu}+\frac{(p_1-p_3)_\mu(p_1-
p_3)_\nu}{m_{D^*}^2}\Big), \nonumber \\
&&{\cal M}_{X_2}^{(d)}=g_{\pi D^*D}g_{X_2 D^*D}\pi^{\mu\alpha}_{1}
\frac{1}{u-m_{D^*}^2}(p_2+p_3)^\nu p_{4\alpha} \nonumber \\
&&\quad\quad~\times\Big(-g_{\mu\nu}+\frac{(p_1-p_4)_\mu(p_1-
p_4)_\nu}{m_{D^*}^2}\Big) \label{Matx_X2pi2}
\end{eqnarray}
for the $1^+$ state and the $2^-$ state, respectively. And the
amplitudes for the process $X\rho \rightarrow \bar{D^*}D$ are

\begin{eqnarray}
&&{\cal M}_{X_1}^{(g)}=-g_{\rho DD} g_{X_1 D^*D}\epsilon^{\mu}_{1}
\epsilon^{\nu}_{2}\epsilon_{3\mu}^*\frac{1}{t-m_D^2}(2p_4-p_2)_\nu,
\nonumber \\
&&{\cal M}_{X_1}^{(h)}=-g_{\rho D^*D^*}g_{X_1 D^*D}\epsilon^{\mu
}_{1}\epsilon^{\alpha}_{2}\epsilon^{*\beta}_{3}\frac{1}{u-m_{D^*}^2}
\nonumber \\
&&\quad\times\Big(-g_{\mu\nu}+\frac{(p_1-p_4)_\mu(p_1-
p_4)_\nu}{m_{D^*}^2}\Big) \nonumber \\
&&\quad\times\Big((2p_3-p_2)_\alpha g^\nu_\beta-(p_3+p_2
)^\nu g_{\alpha\beta}+(2p_2-p_3)_\beta g^\nu_\alpha\Big) \nonumber \\
\label{Matx_X1rho1}
\end{eqnarray}
and

\begin{eqnarray} &&{\cal
M}_{X_2}^{(g)}=-g_{\rho DD} g_{X_2 D^*D}\pi^{\mu\gamma
}_{1}\epsilon^{\nu}_{2}\epsilon_{3\mu}^* \nonumber \\
&&\quad\times\frac{1}{t-m_D^2}(2p_4-p_2)_\nu(p_1-
p_3)_\gamma, \nonumber \\
&&{\cal M}_{X_2}^{(h)}=-g_{\rho D^*D^*}g_{X_2D^*D}\pi^{\mu\gamma
}_{1}\epsilon^{\alpha}_{2}\epsilon^{*\beta}_{3}\frac{1}{u-m_{D^*}^2}
p_{4\gamma} \nonumber \\
&&\quad\times\Big(-g_{\mu\nu}+\frac{(p_1-p_4)_\mu(p_1-
p_4)_\nu}{m_{D^*}^2}\Big) \nonumber \\
&&\quad\times\Big((2p_3-p_2)_\alpha g^\nu_\beta -(p_3+p_2)^\nu
g_{\alpha\beta}+(2p_2-p_3)_\beta
g^\nu_\alpha\Big). \nonumber \\
\label{Matx_X2rho1}
\end{eqnarray}
Finally the amplitudes for the processes $X\rho \rightarrow
D\bar{D}$ and $X\rho \rightarrow D^*\bar{D}^*$ are

\begin{eqnarray} &&{\cal
M}_{X_1}^{(i)}=g_{\psi DD} g_{X_1\psi\rho}\varepsilon^{
\mu\nu\rho\sigma}\epsilon_{1\mu}\epsilon_{2\rho}\frac{1}
{s-m_{\psi}^2}(p_4-p_3)_\alpha p_{1\sigma} \nonumber \\
&&\quad\times\Big(-g_{\nu}^\alpha+\frac{(p_1+p_2)_\nu(p_1+
p_2)^\alpha}{m_{\psi}^2}\Big) \nonumber \\
&&{\cal M}_{X_2}^{(i)}=-g_{\psi DD} g_{X_2\psi\rho}\varepsilon^{
\mu\nu\rho\sigma}\pi_{1\mu\alpha}\epsilon_{2\sigma}\frac{1}
{s-m_{\psi}^2}(p_4-p_3)_\beta \nonumber \\
&&\quad\times (p_1+p_2)_\nu p_{2\rho}\Big(-g^{\alpha\beta}+
\frac{(p_1+p_2)^\alpha(p_1+p_2)^\beta}{m_{\psi}^2}\Big) \nonumber \\
&&\quad+g_{\psi DD} g_{X_2\psi\rho}\varepsilon^{
\mu\nu\rho\sigma}\pi_{1\mu\alpha}\epsilon_2^\alpha\frac{1}
{s-m_{\psi}^2}(p_4-p_3)_\beta \nonumber \\
&&\quad\times (p_1+p_2)_\rho p_{2\nu}\Big(-g_{\sigma}^\beta+
\frac{(p_1+p_2)_\sigma(p_1+p_2)^\beta}{m_{\psi}^2}\Big) \nonumber \\
&&\quad+g_{\psi DD} g_{X_2\psi\rho}'\varepsilon^{
\mu\nu\rho\sigma}\pi_{1\mu\alpha}\epsilon_{2\sigma}\frac{1}
{s-m_{\psi}^2}(p_4-p_3)_\beta \nonumber \\
&&\quad\times (p_1+2p_2)^\alpha p_{1\nu}\Big(-g_{\rho}^\beta
+\frac{(p_1+p_2)_\rho(p_1+ p_2)^\beta}{m_{\psi}^2}\Big), \nonumber \\
\label{Matx_Xrho2}
\end{eqnarray}
and

{\allowdisplaybreaks\begin{eqnarray} &&{\cal
M}_{X_1}^{(j)}=g_{\psi D^*D^*}g_{X_1\psi\rho}\varepsilon^{
\mu\nu\rho\sigma}\epsilon_{1\mu}\epsilon_{2\rho}\epsilon_{3\gamma}^*
\epsilon_{4\beta}^*\frac{1}{s-m_{\psi}^2} p_{1\sigma} \nonumber \\
&&\quad\times \Big((p_3- p_4)_\alpha g^{\gamma\beta}-(2p_3+
p_4)^\beta g_\alpha^\gamma+(p_3+2p_4)^\gamma g_\alpha^{\beta}\Big)
\nonumber \\
&&\quad\times\Big(-g_{\nu}^\alpha+\frac{(p_1+p_2)_\nu(p_1+p_2
)^\alpha}{m_{\psi}^2}\Big) \nonumber \\
&&{\cal M}_{X_2}^{(j)}=g_{\psi
D^*D^*}\varepsilon^{\mu\nu\rho\sigma}\pi_{1\mu\alpha} \epsilon_{
3\gamma}^*\epsilon_{4\delta}^*\frac{1}{s-m_{\psi}^2} \nonumber \\
&&\quad\times \Big((p_3-p_4 )_\beta g^{\gamma\delta}-(2p_3+
p_4)^\delta g_\beta^\gamma+(p_3+2p_4)^\gamma g_\beta^{\delta}
\Big) \nonumber \\
&&\quad\times \bigg[ -g_{X_2\psi\rho}\epsilon_{2
\sigma} \Big(-g^{\alpha\beta}+\frac{(p_1+p_2)^\alpha(p_1+p_2
)^\beta}{m_{\psi}^2}\Big) \nonumber \\
&&\quad\times (p_1+p_2)_\nu p_{2\rho}+g_{X_2\psi\rho}
\epsilon_2^\alpha(p_1+p_2)_\rho p_{2\nu} \nonumber \\
&&\quad\times \Big(-g_{\sigma}^\beta+
\frac{(p_1+p_2)_\sigma(p_1+p_2)^\beta}{m_{\psi}^2}\Big)+g_{X_2\psi\rho}'
\epsilon_{2\sigma} \nonumber \\
&&\quad\times(p_1+2p_2)^\alpha p_{1\nu} \Big(-g_{\rho}^\beta
+\frac{(p_1+p_2)_\rho(p_1+p_2)^\beta}{m_{\psi}^2}\Big)\bigg], \nonumber \\
\label{Matx_Xrho3}
\end{eqnarray}}
respectively, for both the $X_1(3872)$ meson and the $X_2(3872)$
meson.

In the above equations, $p_j$ denotes the momentum of particle
$j$. We choose the convention that particles $1$ and $2$ represent
initial-state mesons, while particles $3$ and $4$ represent
final-state mesons on the left and right sides of the diagrams,
respectively. Here we use the usual Mandelstam variables given by
$s=(p_{1}+p_{2})^{2}$, $t=(p_{1}-p_{3})^{2} $, and
$u=(p_{1}-p_{4})^{2}$. The polarization tensor $\pi^{\mu\nu}$
satisfies the following polarization sum:

{\allowdisplaybreaks\begin{eqnarray}
&&\sum_{pol}\pi^{\mu\nu}\pi^{*\mu'\nu'} \nonumber \\
&&=\frac{1}{2}\Big(-g^{\mu\mu'}+
\frac{k^{\mu}k^{\mu'}}{m_X^2}\Big)\Big(-g^{\nu\nu'}+\frac{k^{\nu}
k^{\nu'}}{m_X^2}\Big) \nonumber \\
&&+\frac{1}{2}\Big(-g^{\mu\nu'}+\frac{k^{\mu}k^{\nu'}}{m_X^2}\Big)
\Big(-g^{\mu'\nu}+\frac{k^{\mu'}k^{\nu}}{m_X^2}\Big) \nonumber \\
&&-\frac{1}{3}\Big(-g^{\mu\nu}+\frac{k^{\mu}k^{\nu}}{m_X^2}\Big)
\Big(-g^{\mu'\nu'}+\frac{k^{\mu'}k^{\nu'}}{m_X^2}\Big) \nonumber \\
&&=\frac{1}{2}\Big(g^{\mu\mu'}g^{\nu\nu'}+g^{\mu\nu'}g^{\mu'\nu}
-g^{\mu\nu}g^{\mu'\nu'}\Big)-\frac{1}{2m_X^2} \nonumber \\
&&\times\Big(g^{\mu\mu'}k^{\nu}k^{\nu'}+g^{\nu\nu'}
k^{\mu}k^{\mu'}+g^{\mu\nu'}k^{\mu'}k^{\nu}+g^{\mu'\nu}k^{\mu}
k^{\nu'}\Big)\nonumber \\
&&+\frac{1}{6}\Big(g^{\mu\nu}+\frac{2}{m_X^2}k^{\mu}k^{\nu}\Big)\Big(
g^{\mu'\nu'}+ \frac{2}{m_X^2}k^{\mu'}k^{\nu'}\Big).
\end{eqnarray} }

In obtaining the full amplitudes we introduce the following form
factors at interaction vertices to prevent the artificial growth
of the tree-level amplitudes with the energy:

\begin{equation}
F(\vec q) =\frac{\Lambda^2}{\Lambda^2 +{\vec q}^2},
\end{equation}
where ${\vec q}^2$ is the squared three-momentum transfer for $t$
and $u$ channels and the squared three-momentum of the incoming
particles for $s$ channel taken in the center-of-mass frame. For
the cutoff parameter $\Lambda$, we use $\Lambda=2.0$ GeV. The
final isospin- and spin-averaged cross section is given by

\begin{equation}
\sigma=\frac{1}{64\pi^2 s g_1g_2}\frac{|\vec p_f|}{|\vec p_i|}\int
d\Omega\overline{|\mathcal{M}|^2}F^4, \label{sigma}
\end{equation}
with $g_1$ and $g_2$ being the degeneracy factors of the initial 1
and 2 particles, $(2I_1+1)(2S_1+1)$ and  $(2I_2+1)(2S_2+1)$,
respectively. We denote by  $\overline{|\mathcal{M}|^2}$  the
squared amplitude of all processes in Eq. (\ref{amplitudes})
obtained by summing over the isospins and spins of both the
initial and final particles after killing all unphysical terms
satisfying $p_{1\mu}\pi_1^{\mu\nu}=0$,
$p_{2\mu}\varepsilon_2^{\mu}=0$ and so on. In evaluating
$\overline{|\mathcal{M}|^2}$ we only consider the $X(3872)$ meson
interacting with $D^0$ and $\bar{D}^{0*}$ (or $\bar{D}^0$ and
$D^{0*}$ ) since the $X(3872)$ meson does not have its isospin
partner. In Eq. (\ref{sigma}), $|\vec p_i|$ and $|\vec p_f|$
represent the three-momenta of the initial and final particles in
the center of mass frame. For coupling constants, we use $g_{\rho
DD}=g_{\rho D^*D^*}$=2.52, $g_{\psi DD}=g_{\psi D^*D^*}$=7.64 from
\cite{Lin:1999ad}, and $g_{\pi D^*D}$=6.3 from the decay width of
$D^*$ meson \cite{Ahmed:2001xc}. The strong-coupling constants for
the $X(3872)$ meson have been taken from Table II in Ref.
\cite{Brazzi:2011fq}, and those are summarized in Table
\ref{Xcouplings}.

\begin{table}[!h]
\caption{The strong-coupling constants for the $X(3872)$ meson
\cite{Brazzi:2011fq}. } \label{Xcouplings}
\begin{center}
\begin{tabular}{ccc}
\hline \hline
& J$^p=1^+$ & J$^p=2^-$  \\
\hline
$g_{X_JD^*D}$ & 3.5 $\pm$ 0.7 GeV & 189 $\pm$ 36 \\
$g_{X_J\psi\rho}$ & 0.14 $\pm$ 0.03 & -0.29 $\pm$ 0.08 GeV$^{-1}$ \\
$g_{X_J\psi\rho}'$ &  & 0.28 $\pm$ 0.09 GeV$^{-1}$ \\
\hline \hline
\end{tabular}
\end{center}
\end{table}

\begin{figure}[!h]
\begin{center}
\includegraphics[width=0.50\textwidth]{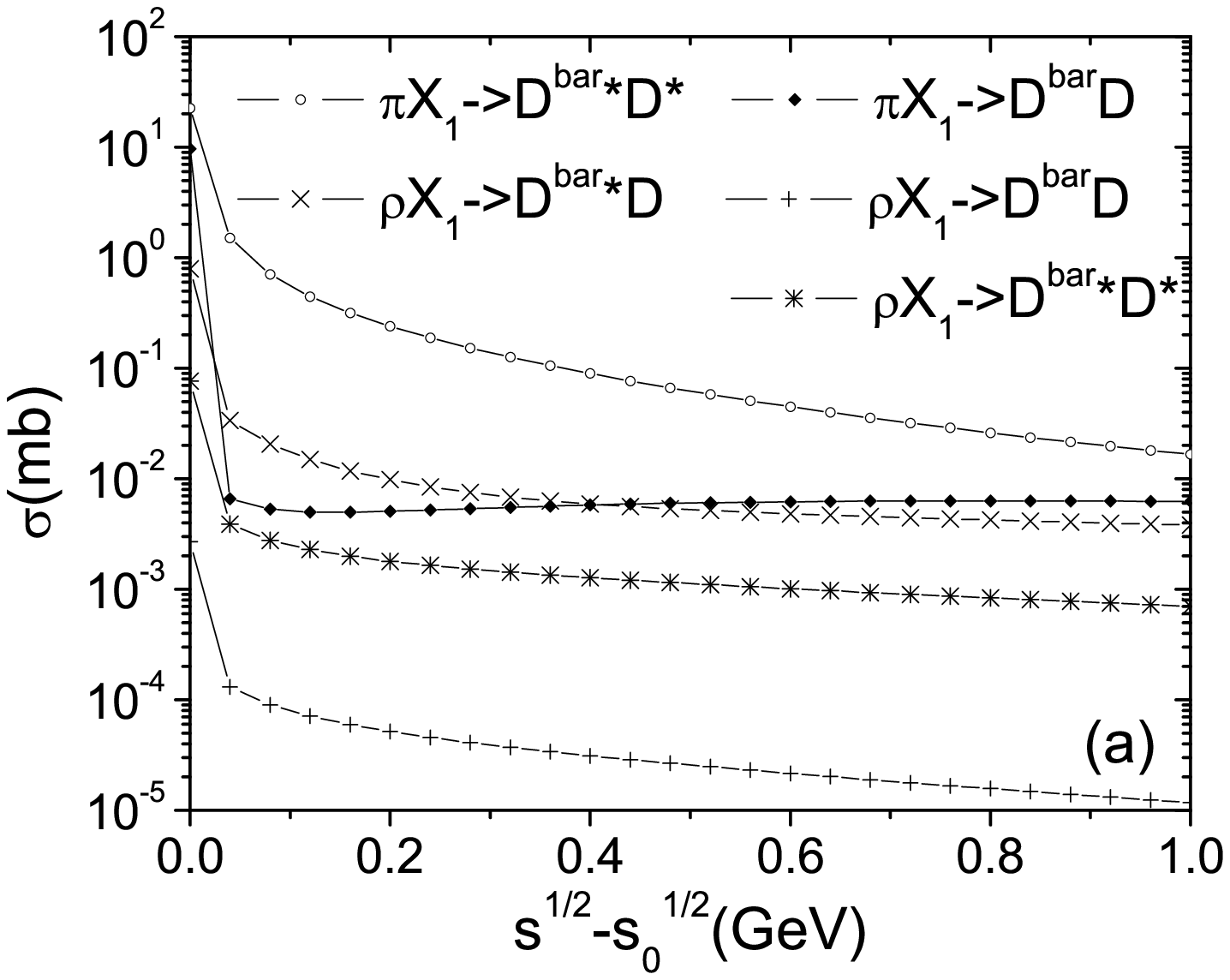}
\includegraphics[width=0.50\textwidth]{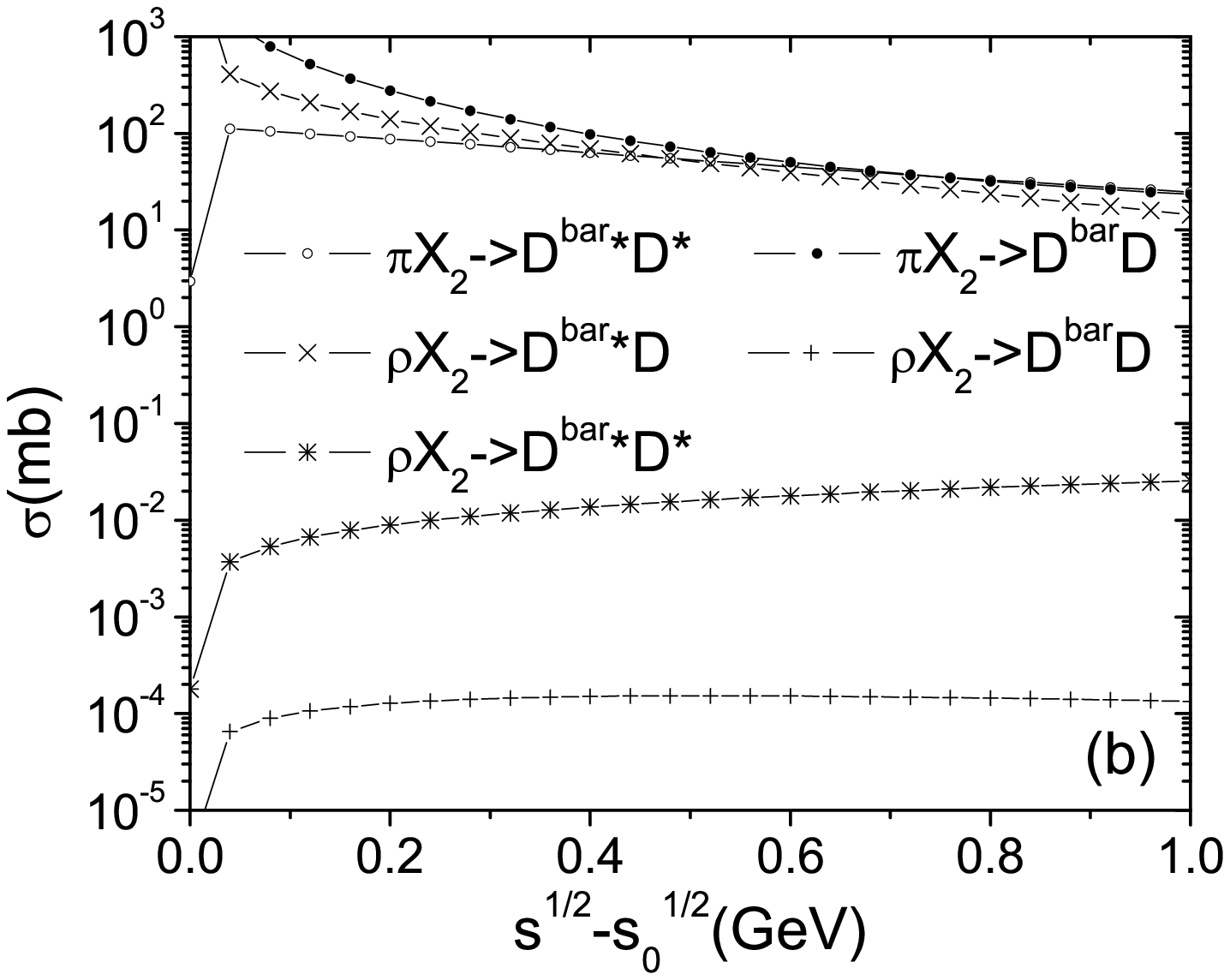}
\end{center}
\caption{Cross sections for the absorption of (a) a $X_1(3872)$
meson and (b) a $X_2(3872)$ meson by $\pi$ and $\rho$ mesons via
processes X$\pi\rightarrow \bar{D}^*D^*$, X$\pi\rightarrow
\bar{D}D$, X$\rho\rightarrow\bar{D}^*D$, X$\rho\rightarrow
D\bar{D}$, and X$\rho\rightarrow D^*\bar{D}^*$. }
\label{X1X2sigma}
\end{figure}
In Fig. \ref{X1X2sigma}, we show the cross sections for the
absorption of both a $X_1(3872)$ meson and a $X_2$(3872) meson by
pions and $\rho$ mesons via processes $X\pi\rightarrow
\bar{D}^*D^*$, $X\pi\rightarrow \bar{D}D$,
$X\rho\rightarrow\bar{D}^*D$, $X\rho\rightarrow D\bar{D}$, and
$X\rho\rightarrow D^*\bar{D}^*$ as functions of the total
center-of-mass energy $s^{1/2}$ above the threshold energy
$s_0^{1/2}$ of each process. We see in Fig. \ref{X1X2sigma} (a)
that there exists a peak near the threshold energy for the
endothermic process $X_1\pi\rightarrow \bar{D}^*D^*$ ($s_0^{1/2}$=
4013.96 MeV), while the cross sections for the other exothermic
processes $X_1\pi\rightarrow \bar{D}D$ ($s_0^{1/2}$= 4009.72 MeV),
$X_1\rho\rightarrow \bar{D}^*D$, $X_1\rho\rightarrow D\bar{D}$,
and $X_1\rho\rightarrow D^*\bar{D}^*$ ($s_0^{1/2}$= 4647.17 MeV)
become infinite near the threshold. The same general behaviors as
these can be found for the spin-2 $X(3872)$ meson in Fig.
\ref{X1X2sigma}(b). For the endothermic process $X_2\pi\rightarrow
\bar{D}^*D^*$, however, there is a gradual decrease after a strong
rise near the threshold. It is also noticeable that there exists a
sharp dip very near the threshold and a gradual increase and
decrease afterwards for the exothermic processes
$X_2\rho\rightarrow D\bar{D}$ and $X_2\rho\rightarrow
D^*\bar{D}^*$, which is similar to that shown in Ref.
\cite{Brazzi:2011fq} for the dissociation cross section of
$J/\psi$ into the open-charm meson mediated by the spin-2
$X(3872)$ meson. These differences are due to two different
interaction mechanisms originating from two possible spin quantum
numbers. The additional derivative in the interaction Lagrangian
for the spin-2 $X(3872)$ meson causes the completely different
energy dependence on the cross sections especially for the
processes $X\rho\rightarrow D\bar{D}$ and $X\rho\rightarrow
D^*\bar{D}^*$.

We also clearly see in Fig. \ref{X1X2sigma} that the absorption
cross sections for the processes $X\pi\rightarrow \bar{D}^*D^*$,
$X\pi\rightarrow \bar{D}D$, and $X\rho\rightarrow\bar{D}^*D$ are
much bigger when the spin of $X(3872)$ mesons is 2 than when it is
1. This is largely attributed to the large strong-coupling
constant used to evaluate the cross sections for the processes
having $X_2D^*D$ interactions. For the processes involving the
$\rho$ meson, the additional derivative in the interaction
Lagrangian for the spin-2 $X(3872)$ meson causes 10-40 times
bigger cross sections in the processes $X_2\rho\rightarrow
D\bar{D}$ and $X_2\rho\rightarrow D^*\bar{D}^*$ compared to those
in the processes $X_1\rho\rightarrow D\bar{D}$ and
$X_1\rho\rightarrow D^*\bar{D}^*$ as shown in Fig. \ref{X1X2sigma}
when the strong couplings $g_{X_1\psi\rho}$, $g_{X_2\psi\rho}$,
and $g_{X_2\psi\rho}'$ have been used. In the processes
$X\pi\rightarrow \bar{D}^*D^*$, $X\pi\rightarrow \bar{D}D$, and
$X\rho\rightarrow\bar{D}^*D$, however, the additional derivative
brings out roughly $m_D \sim 1.9$ GeV. This factor will multiply
the already large coupling constant $g_{X_2D^*D}$ given in Table
\ref{Xcouplings}, making the effective coupling strength ( $\sim
189\times 1.9$ GeV ) much larger compared to  $g_{X_1D^*D} = 3.5$
GeV. This explains the bigger cross sections for the $X_2(3872)$
meson than those for the $X_1(3872)$ meson in Fig.
\ref{X1X2sigma}.

Both strong coupling constants $g_{X_1D^*D}$ and $g_{X_2D^*D}$
were obtained from \textit{one} experimental measurement using
\textit{two} different spin possibilities \cite{Brazzi:2011fq}. As
was already pointed out, however, the analysis of the $X(3872)$
meson decaying to $\bar{D}^0D^{*0}$ disfavors the $2^-$ quantum
number because of the angular momentum barrier in its
near-threshold decay \cite{Gokhroo:2006bt,Aubert:2007rva}. The $D$
meson should have the relative angular momentum in order to be
able to interact with the spin-2 $X(3872)$ meson to satisfy the
angular momentum conservation. Therefore, the coupling constant
$g_{X_2D^*D}$ has to be large to compensate for the angular
momentum suppression near threshold.

Nevertheless, it is still possible to get a smaller strong
coupling constant $g_{X_2D^*D}$ when the $X(3872)$ meson mass
increases slightly. In Appendix A, we have investigated the origin
of the big strong-coupling constant $g_{X_2D^*D}$, and have found
that it is very sensitive to the variation of the $X(3872)$ meson
mass. Varying the mass within the experimental uncertainty, we
estimate that $g_{X_2D^*D}$ could be reduced by a factor of
$\sqrt{3}$ and the cross sections of the spin-2 $X(3872)$ meson
evaluated in Fig. \ref{X1X2sigma} by a factor of 3.

The bigger cross sections for the spin-2 $X(3872)$ meson are
contrary to naive expectations. It is expected that the size of
the bag containing four quarks should be at least bigger than that
of the bag having two quarks. In the simple bag model the size of
the bag increases with the number of quarks inside the bag as $R
\propto N_q^{1/4}$ \cite{Fewbody}. Since the cross section depends
on the size of the hadron in general, we expect the cross section
of the $X(3872)$ meson composed of four quarks to be bigger than
that of the $X(3872)$ meson made up of two quarks. However, we see
here only the effects from the interaction mechanism caused by two
different spins since the interaction Lagrangians are blind to the
size of the $X(3872)$ meson.

\section{Time evolution of the $X(3872)$ meson abundance in hadronic matter}

Using the cross sections evaluated in the previous section we now
consider the time evolution of the $X(3872)$ meson abundance in
hadronic matter. We build the evolution equation consisting of the
densities and abundances for hadrons participating in all
processes shown in Fig. \ref{Diagrams}: $\pi$, $\rho$, $D^*$, and
$D$ mesons.

\begin{widetext}

\begin{figure}[!h]
\begin{center}
\includegraphics[width=0.49\textwidth]{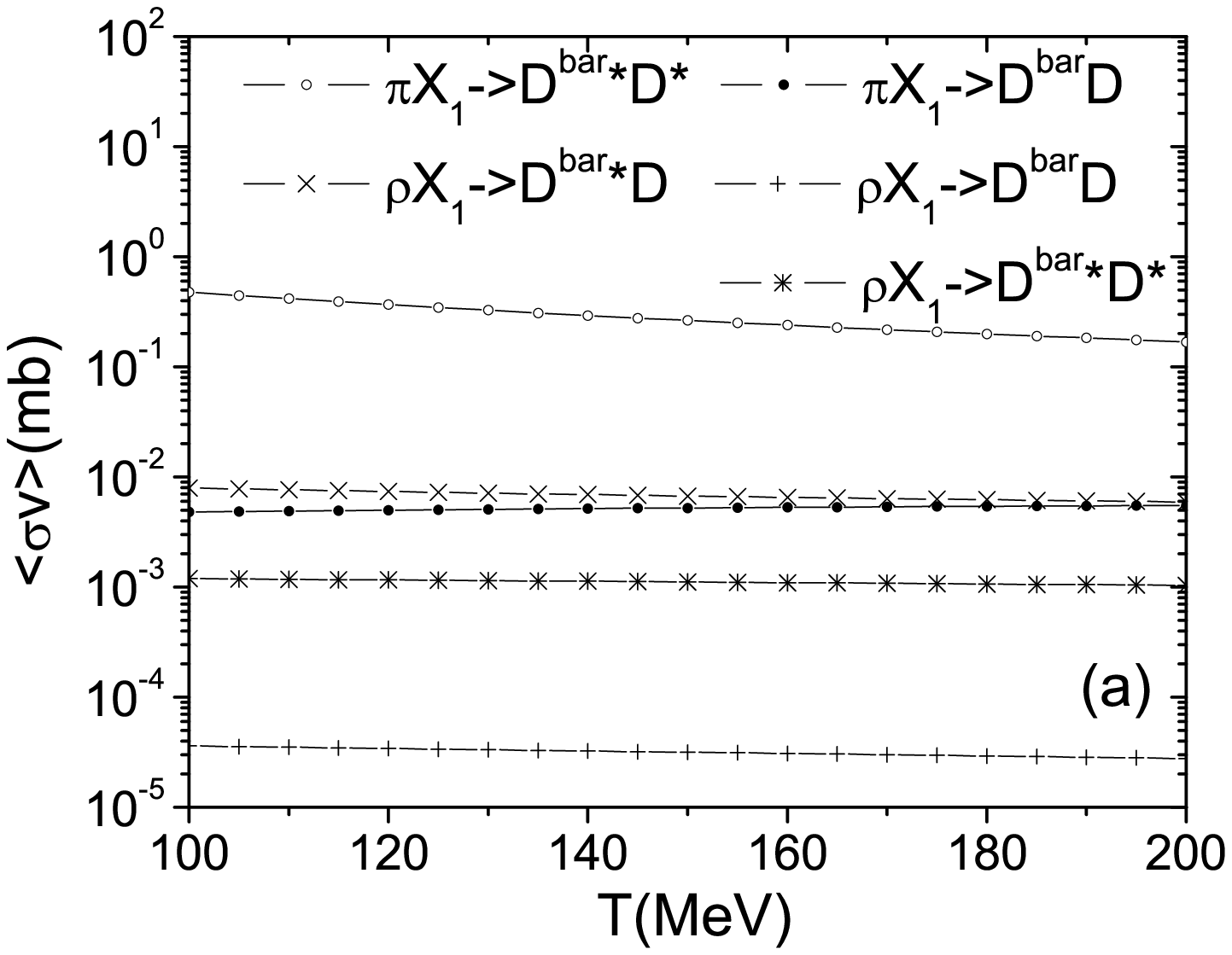}
\includegraphics[width=0.49\textwidth]{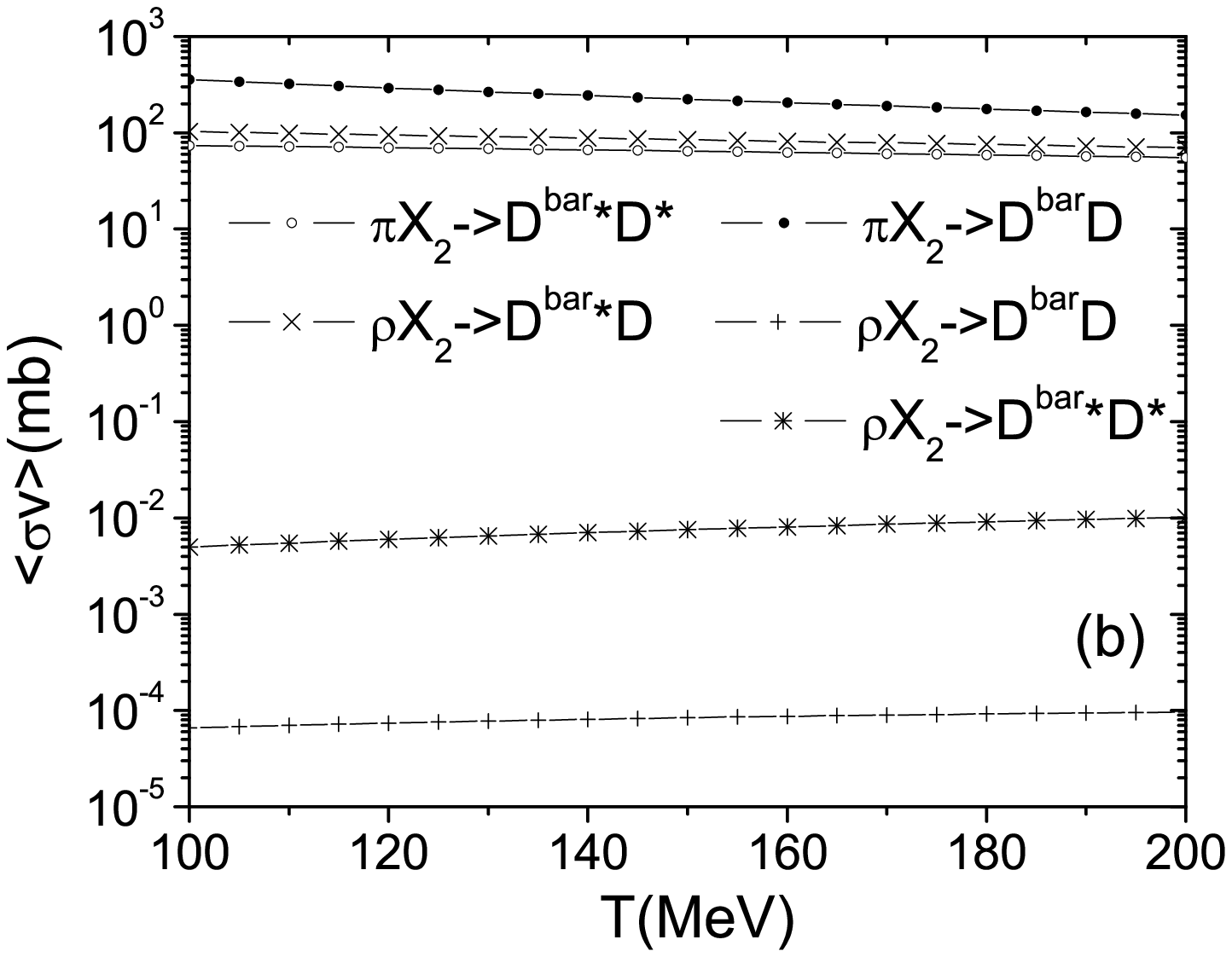}
\includegraphics[width=0.49\textwidth]{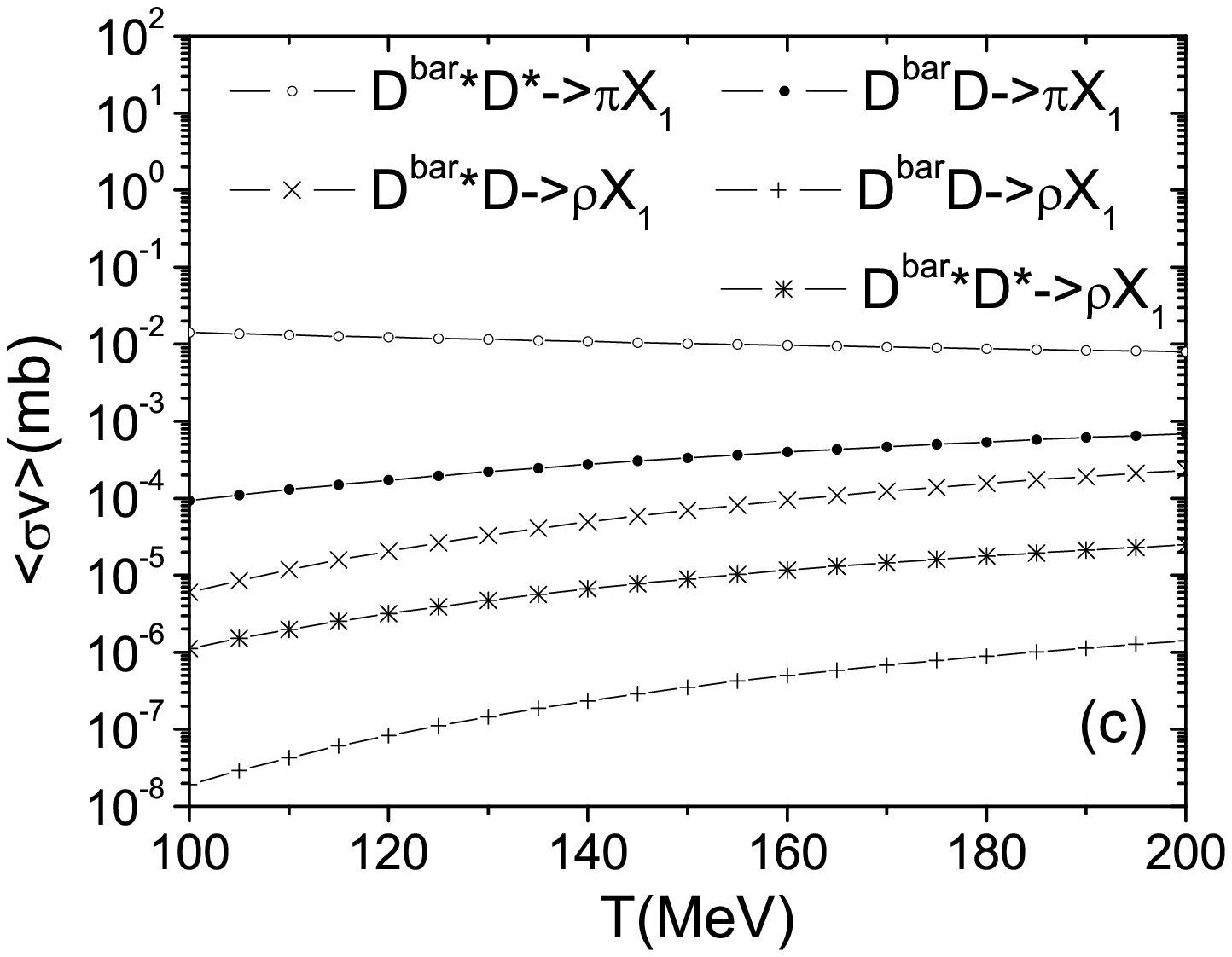}
\includegraphics[width=0.49\textwidth]{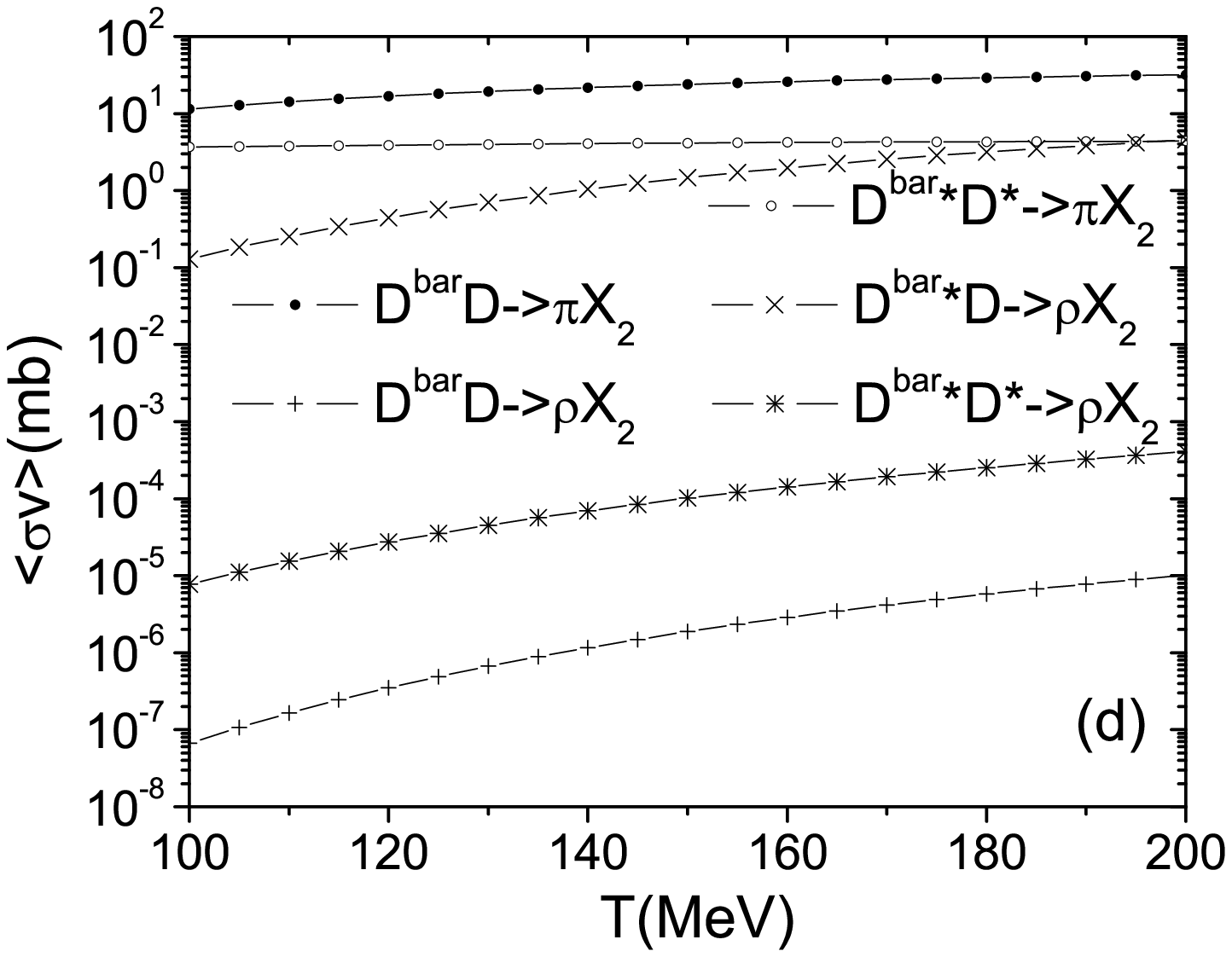}
\end{center}
\caption{Thermally averaged cross sections for the absorption of
(a) a $X_1(3872)$ meson and (b) a $X_2(3872)$ meson by pions and
$\rho$ mesons via processes $X\pi\rightarrow \bar{D}^*D^*$,
$X\pi\rightarrow \bar{D}D$, $X\rho\rightarrow\bar{D}^*D$,
$X\rho\rightarrow D\bar{D}$, and $X\rho\rightarrow D^*\bar{D}^*$
and their inverse processes $\bar{D}^*D^*\rightarrow$$X\pi$,
$\bar{D}D\rightarrow$$X\pi$, $\bar{D}^*D\rightarrow$$X\rho$,
$D\bar{D}\rightarrow$$X\rho$, and $D^*\bar{D}^*\rightarrow$$X\rho$
for (c) a $X_1(3872)$ meson and (d) a $X_2(3872)$ meson.}
\label{X1X2thsigma}
\end{figure}

\end{widetext}

\begin{eqnarray}
&&\frac{dN_X(\tau)}{d\tau}=R_{QGP}(\tau)+\sum_{l,c,c'}\bigg(\langle
\sigma_{cc'\to lX}v_{cc'}\rangle n_{c}(\tau)N_{c'}(\tau) \nonumber \\
&&\qquad\qquad-\langle\sigma_{lX\to cc'}v_{lX}\rangle n_l(\tau)
N_X(\tau) \bigg), \label{rate}
\end{eqnarray}
where $n_l(\tau)$ and $n_c(\tau)$ are, respectively, the density
of a light meson such as a pion or a $\rho$ meson and the density
of a charmed meson in the hadronic matter at proper time $\tau$,
whereas $N_{c'}(\tau)$ is the abundance of the other charmed meson
in each process shown in Fig. \ref{Diagrams} at proper time
$\tau$. $n_l(\tau)$, $n_c(\tau)$ and $N_{c'}(\tau)$ are calculated
from Eq. (\ref{Stat}) by assuming that light mesons and charmed
mesons are in equilibrium and vary in time through the temperature
profile introduced below, Eq. (\ref{profiles}). In the above rate
equation, Eq. (\ref{rate}), $\left\langle \sigma _{ab\rightarrow
cd}v_{ab}\right\rangle$ is the cross section averaged over the
thermal distribution for initial two particles in a two-body
process $ab\to cd$ given by \cite{Koch:1986ud}

\begin{eqnarray}
&&\left\langle \sigma _{ab\rightarrow cd}v_{ab}\right\rangle
\nonumber \\
&&=\frac{\int d^{3}\mathbf{p}_{a}d^{3}\mathbf{p}_{b}f_{a}
(\mathbf{p}_{a})f_{b}(\mathbf{p}_{b})\sigma_{ab\to cd}v_{ab}}{\int
d^{3}\mathbf{p}_{a}d^{3}\mathbf{p}_{b}f_{a}(\mathbf{p}_{a})
f_{b}(\mathbf{p}_{b})} \nonumber\\
&&=\frac{1}{4\alpha^2_a K_2(\alpha_a)\alpha^2_b K_2(\alpha_b)}
\int^\infty_{z_0}dzK_1(z)\sigma(s=z^2 T^2) \nonumber \\
&&\times[z^2-(\alpha_a+\alpha_b)^2][z^2-(\alpha_a-\alpha_b)^2],
\end{eqnarray}
with $\alpha_i=m_i/T$, $z_0=\mathrm{max}(\alpha_a+\alpha_b,
\alpha_c +\alpha_d)$, $K_1$ and $K_2$ being the modified Bessel
function of the first and second kind, respectively and $v_{ab}$
denoting the relative velocity of the initial two interacting
particles $a$ and $b$, $v_{ab}=\sqrt{(p_a\cdot
p_b)^2-m^2_am^2_b}/(E_aE_b)$.

The $X(3872)$ meson abundance at proper time $\tau$, $N_X(\tau)$,
depends on both the dissociation rate such as $X\pi\rightarrow
\bar{D}^*D^*$, $X\pi\rightarrow \bar{D}D$,
$X\rho\rightarrow\bar{D}^*D$, $X\rho\rightarrow D\bar{D}$, and
$X\rho\rightarrow D^*\bar{D}^*$ and the production rate through
the inverse processes, $\bar{D}^*D^*\rightarrow$$X\pi$,
$\bar{D}D\rightarrow$$X\pi$, $\bar{D}^*D\rightarrow$$X\rho$,
$D\bar{D}\rightarrow$$X\rho$, and
$D^*\bar{D}^*\rightarrow$$X\rho$. We use the detailed balance
relations based on the results for the forward processes shown in
Fig. \ref{X1X2sigma} in evaluating the thermally averaged cross
sections of the inverse processes. The results are shown in Fig.
\ref{X1X2thsigma}.

In Eq. (\ref{rate}), $n_l(\tau)$, $n_c(\tau)$ and $N_{c'}(\tau)$
varies in time through the temperature profile developed to
describe the dynamics of relativistic heavy ion collisions. We use
the schematic model based on the boost invariant Bjorken picture
with an accelerated transverse expansion \cite{Chen:2003tn,
Chen:2007zp}. The system of the quark-gluon plasma of its final
transverse size $R_C$ at the chemical freeze-out time $\tau_C$
expands with its transverse velocity $v_C$ and transverse
acceleration $a_C$. The temperature of the system is maintained
with a constant temperature $T_C$ until the end of the mixed phase
at $\tau_H$, and decreases afterwards to the kinetic freeze-out
temperature $T_F$. The volume and temperature profiles as a
function of the proper time $\tau$ are as follows:

\begin{eqnarray}
&& V(\tau)=\pi[R_C+v_C(\tau-\tau_C)+a_C/2(\tau-\tau_C)^2]^2\tau
c, \nonumber \\
&& T(\tau)=T_C-(T_H-T_F)\bigg(\frac{\tau-\tau_H}{\tau_F-\tau_H}
\bigg)^{4/5}, \label{profiles}
\end{eqnarray}
with $T_H$ and $\tau_F$ being the hadronization temperature, and
the freeze-out time, respectively. The values used in Eq.
(\ref{profiles}) are summarized in Table \ref{data}.

\begin{table}[!h]
\caption{Values for the volume and temperature profiles in the
schematic model Eq. (\ref{profiles}). } \label{data}
\begin{center}
\begin{tabular}{ccc}
\hline \hline
& Temp.(MeV) & Time (fm/c) \\
\hline
$R_C=8.0$ fm & $T_C=175$  & $\tau_C=5.0$  \\
$v_C=0.4 c$ & $T_H=175$  & $\tau_H=7.5$  \\
$a_C=0.02 c^2/$fm & $T_F=125$  & $\tau_F=17.3$  \\
\hline \hline
\end{tabular}
\end{center}
\end{table}

In solving Eq. (\ref{rate}) we have assumed that the total number
of charm quarks in charmed hadrons is conserved during the
evolution of the hadronic matter. It has been discussed that
chances for charmed mesons to be produced and destroyed in the
hadronic matter are very small because of their small production
and annihilation cross sections \cite{Lin:1999ad, Oh:2000qr,
Lin:1999ve}. The light mesons are assumed to be in equilibrium
with the medium and the total number of the pion is set to 926 at
freeze-out \cite{Chen:2003tn} and that of the $\rho$ meson to 68
after considering the contributions from the decays of resonances.
To take into account the effect of the production of the $X(3872)$
meson through hadronization from the quark-gluon plasma, we
include the term $R_{QGP}(\tau)$ \cite{Chen:2007zp} given by

\begin{eqnarray}
R_{QGP}(\tau)=\left\{ \begin{array}{ll} N_X^0/(\tau_H-\tau_C), &
\tau_C < \tau < \tau_H \\
0, & \textrm{otherwise} \end{array}\right. \label{RQGP}
\end{eqnarray}
with $N_X^0$ being the total number of the $X(3872)$ meson in
Table \ref{yields} produced from quark-gluon plasma either by the
two-quark coalescence or by the four-quark coalescence. We assume
here that the volume of the quark-gluon plasma decreases linearly
during the phase-transition time $\tau_H-\tau_C=2.5$ fm/$c$
whereas that of the hadron gas increases with a rate enough to
occupy both the decreased volume of the quark-gluon plasma and the
newly increased volume of the entire system by the expansion.

\begin{figure}[!h]
\begin{center}
\includegraphics[width=0.50\textwidth]{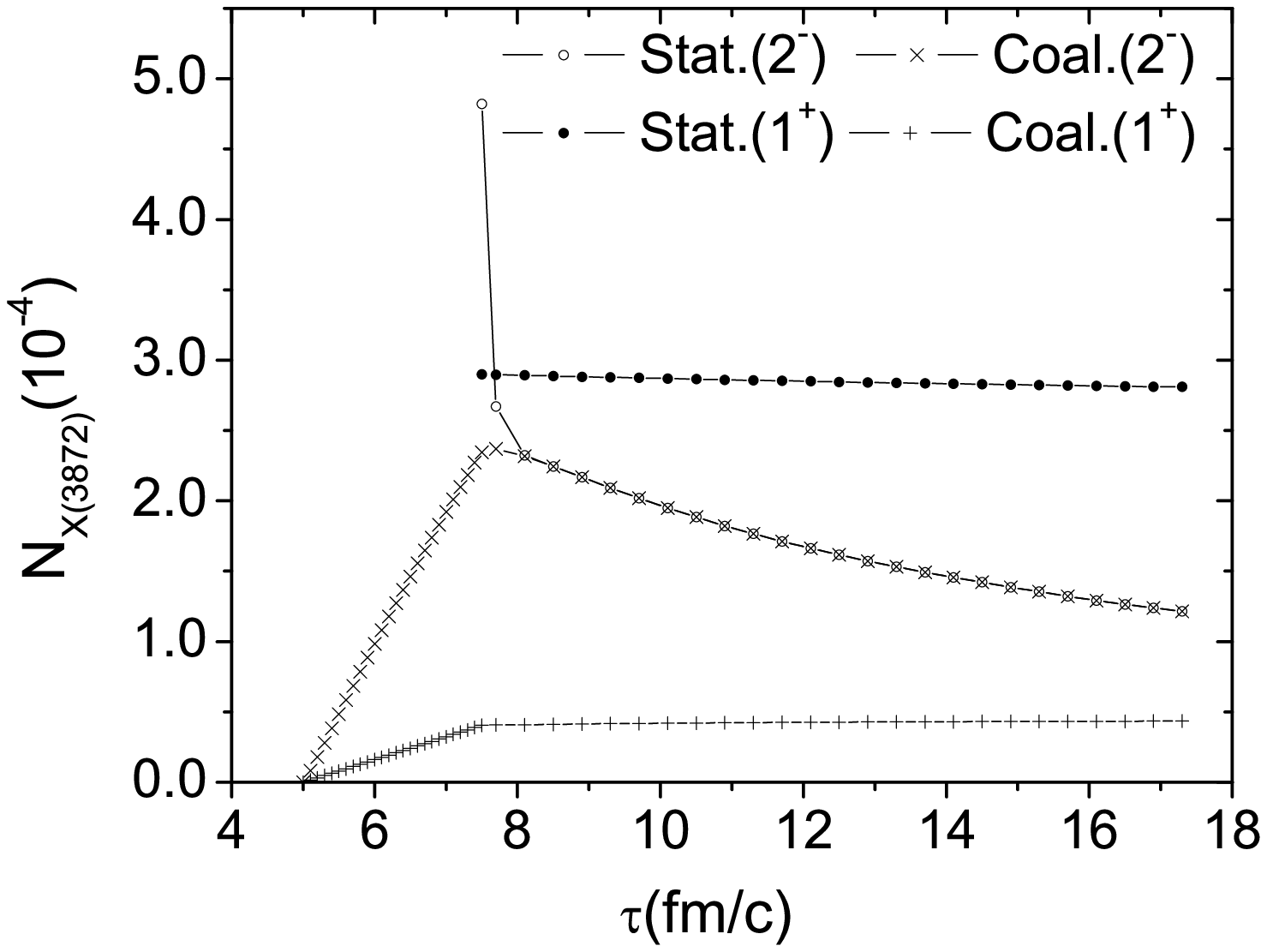}
\end{center}
\caption{Time evolution of $X(3872)$ meson abundances in central
Au-Au collisions at $\sqrt{s_{NN}}$ = 200 GeV for different states
produced from the quark-gluon plasma.} \label{Evol}
\end{figure}
In Fig. \ref{Evol}, we show the abundances of the $X(3872)$ meson
as a function of the proper time for different states produced
from the quark-gluon plasma in central Au-Au collisions at
$\sqrt{s_{NN}}$ = 200 GeV. Since the scattering cross sections for
the $1^{+}$ state $X_1(3872)$ meson are so small as shown in Fig.
\ref{X1X2sigma}, the abundance obtained by the four-quark
coalescence increases very slightly to $4.3\times10^{-5}$, while
the expectation from the statistical model decreases also very
slightly to $2.8\times10^{-4}$. However, due to the large
scattering cross sections for the $2^{-}$ state $X(3872)$ meson,
the $X(3872)$ meson in the normal $c\bar{c}$ state with $d$-wave
has more chances to interact with light mesons in the hadronic
evolution, and therefore the abundance decreases fast to
1.2$\times10^{-4}$. The thermal model expectation for the spin-2
$X(3872)$ meson also decreases rapidly, follows the evolution of
the coalescence model abundance, and evolutes afterward together.
The final ratio of the abundance for the $X_2(3872)$ meson over
that for the $X_1(3872)$ meson both in the coalescence model is
expected to be $\sim$ 2.8 at the kinetic freeze-out.

Based on the above investigation about the time evolution of the
$X(3872)$ meson, we can further consider the possibility of
producing a hadronic molecular state of the spin-1 $X(3872)$
meson. If the state is a hadronic molecule composed of
$D^0\bar{D}^{*0}(\bar{D}^0D^{*0})$ in $s$-wave, it will be
dominantly produced at the end of the hadronic phase through
hadronic coalescence. The production in the hadronic phase through
the two body hadronic interaction would be very small. For such
process to be possible, the inverse processes like
$\bar{D}^*D^*\rightarrow$$X\pi$, $\bar{D}D\rightarrow$$X\pi$,
$\bar{D}^*D\rightarrow$$X\rho$, $D\bar{D}\rightarrow$$X\rho$, and
$D^*\bar{D}^*\rightarrow$$X\rho$ should prevail the forward
processes. We have already seen, however, in Fig.
\ref{X1X2thsigma} that the thermally averaged cross sections of
the inverse processes are smaller than those of the forward
processes. However, this does not mean that the $X(3872)$ meson
can not be produced in the hadronic phase. In fact, hadronic
coalescence will continue to occur but the total absorption should
be very large since the size of the loosely bound hadronic
molecule is thought to be much bigger than that of the compact
tetraquark state with the same spin.

\begin{figure}[!h]
\begin{center}
\includegraphics[width=0.50\textwidth]{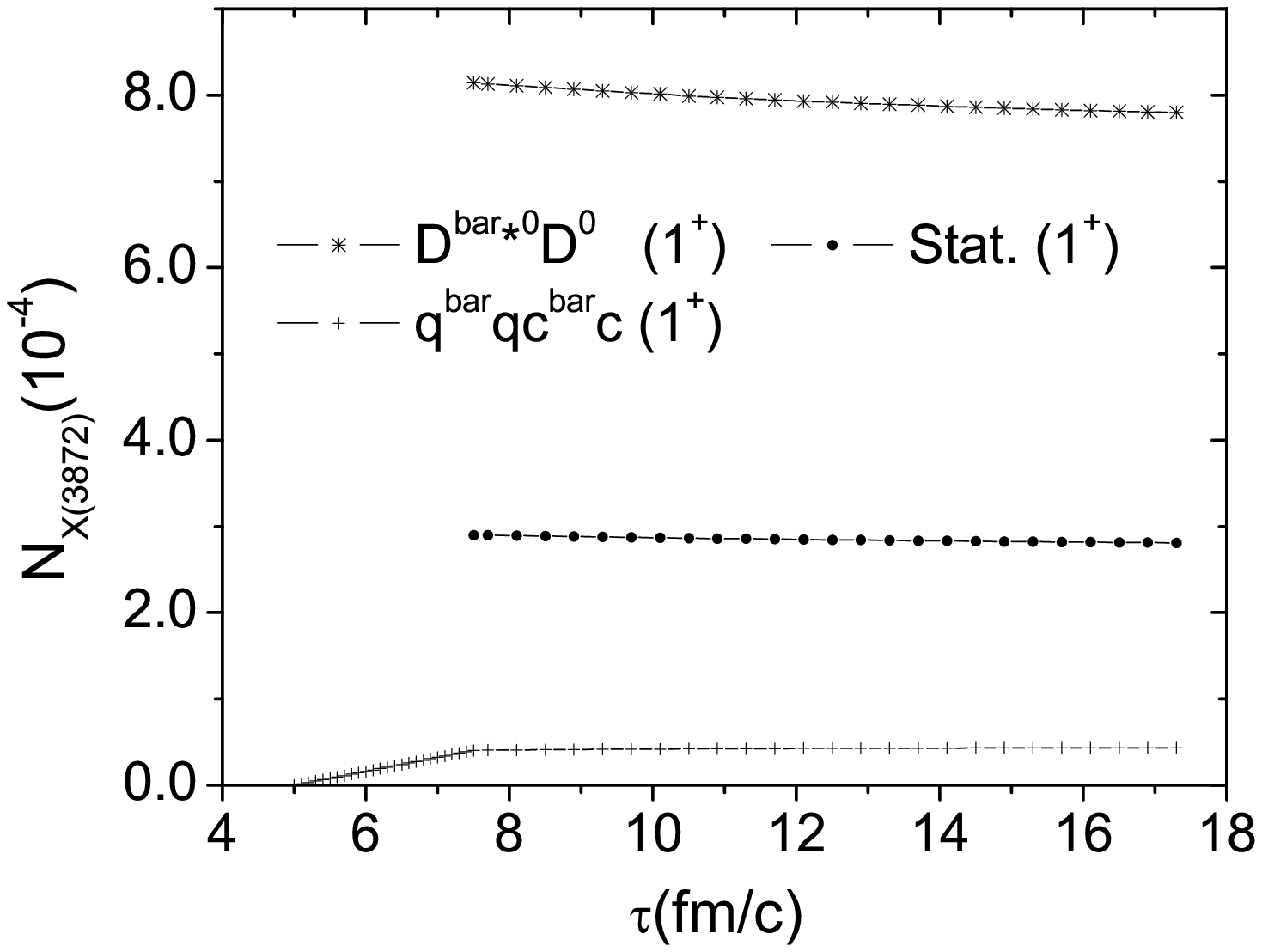}
\end{center}
\caption{The possible time evolution of the hadronic molecular
state of the $X(3872)$ meson produced during the hadronic stage in
central Au-Au collisions at $\sqrt{s_{NN}}$ = 200 GeV when
$X(3872)$ is assumed to be a $1^{+}$ state. } \label{MolEvol}
\end{figure}

With this in mind we can estimate the number of $X(3872)$ mesons
in the hadronic phase by solving the rate equation, Eq.
(\ref{rate}) for the hadronic molecular state backwards in time
from the final yield of $7.8\times10^{-4}$ \cite{Cho:2010db,
Cho:2011ew} calculated through the hadron coalescence at the end
of the hadronic stage. As shown in Fig. \ref{MolEvol}, when the
hadronic molecular state of the $X_1(3872)$ meson is produced
sometime during the hadronic stage by the hadron coalescence, the
number of $X(3872)$ mesons is expected to be in the range between
7.8$\times10^{-4}$ and 8.1$\times10^{-4}$. Owing to the small
cross sections evaluated in this work, which are blind to the size
of the hadron, the yield decreases slightly during the hadronic
stage, finally resulting to the ratio $\sim 18$ between the
hadronic molecular state and the tetraquark state at the kinetic
freeze-out.

The discussion on the production yield for a molecular state of
the $X(3872)$ meson is based on the assumption that it is a
$D^0\bar{D}^{*0}$ state. However, it has been shown that charged
components of $D$ and $D^*$ mesons also play an essential role in
explaining the branching ratio of the $X(3872)$ meson decaying to
$\omega$ and $\rho$ mesons \cite{Gamermann:2009fv,
Gamermann:2009uq, Aceti:2012cb}. If we take into account a linear
combination of $D$ and $D^*$ mesons for the $X(3872)$ meson,
$|X(3872)\rangle=
1/\sqrt{2}(|D^0\bar{D}^{*0}\rangle+|D^+D^{*-}\rangle)$, then we
have to evaluate the average production yields coming from both
$D^0\bar{D}^{*0}$ and $D^+D^{*-}$. Thus, the production yield
would be the same since the numbers of $D$ or $D^*$ mesons are
independent of their charges.

In this analysis, we have used the phenomenological model, Eq.
(\ref{profiles}), assuming the first order phase transition at
hadronization, which is not a true situation in heavy ion
collisions experiment at RHIC top energy. The transition is a
crossover rather than a first order \cite{Gyulassy:2004zy}. We do
not expect, however, that taking the crossover phase transition
into our consideration affects significantly the time evolution of
$X(3872)$ meson abundance during the hadronic stage, since it has
been initiated from the production yield at the hadronization
temperature as shown in Fig. \ref{Evol}. We focus on the
production of the $X(3872)$ meson from a quark-gluon plasma
through coalescence during the crossover transition.

Assuming that the coalescence production of the $X(3872)$ meson
continuously takes place at all temperatures during the crossover
transition, we investigate the explicit temperature-dependent
production of the $X(3872)$ meson using the coalescence formula
obtained from the overlap between the density matrix of the
constituents and the Wigner function of the $X(3872)$ meson. We
note that the yield of the spin-2 $X(3872)$ meson by two-quark
coalescence is $\propto T^2/(1+CT)^3$, whereas that of the spin-1
$X(3872)$ meson by four-quark coalescence is $\propto 1/(1+CT)^3$
with a same constant $C$ \cite{Cho:2011ew}; as the temperature
decreases, the production rate for the spin-1 $X(3872)$ meson
increases but that of the spin-2 $X(3872)$ meson decreases. Hence,
we find that the production of the $X(3872)$ meson through
coalescence during the transition is also dependent on its spin.
We expect the explicit temperature-dependent production rate of
the $X(3872)$ meson, together with the changes in both the number
of constituent quarks and the volume of the quark-gluon plasma
participating in the production of $X(3872)$ mesons during the
crossover transition, to modify the simple linear production of
the $X(3872)$ meson in the first order transition shown in Fig.
\ref{Evol}. The clear picture about hadron production during the
crossover transition is currently not available and needs further
study.

\section{Conclusion}

We have studied the hadronic effects on the $X(3872)$ meson
abundance in heavy ion collisions using one meson exchange model
with the effective Lagrangian. To take into account the effects
due to two different spin possibilities of the $X(3872)$ meson, we
have evaluated two absorption cross sections for both $J^p=1^+$
and $2^-$ states of the $X(3872)$ meson by pions and rho mesons
during the hadronic stage of heavy ion collisions. We have found
that the absorption cross sections and their thermal averages are
strongly dependent on the structure and the quantum number of the
$X(3872)$ meson; the energy dependence of the cross sections is
quite different for two spin states as shown in Fig.
\ref{X1X2sigma}, and the cross sections are much bigger for a
$2^-$ state than those for a $1^+$ state. Therefore, it is
expected that the spin-2 $X(3872)$ meson can be absorbed by light
mesons much more easily than the spin-1 $X(3872)$ meson.

We have further investigated the time evolution of the abundances
for two possible quantum number states of the $X(3872)$ meson. We
have found that the variation of the $X(3872)$ meson abundance
during the expansion of the hadronic matter is also strongly
affected by the quantum number of the $X(3872)$ meson; the
$X_1(3872)$ meson abundance slightly changes to
4.3$\times10^{-5}$, while the $X_2(3872)$ meson abundance varies
significantly to 1.2$\times10^{-4}$, leading to the final
abundance ratio $\sim$ 2.8 between $X_2(3872)$ and $X_1(3872)$
mesons at the kinetic freeze-out. We therefore suggest that
studying the abundance of the $X(3872)$ meson in relativistic
heavy ion collisions provides a chance to infer its quantum number
as well as its structure.

We have also considered the hadronic molecular state
$D^0\bar{D}^{*0}(\bar{D}^0D^{*0})$ possibly produced from
$D^0(\bar{D^0})$ and $\bar{D}^{*0}(D^{*0})$ sometime during the
hadronic stage. The abundance is expected to be in the range
between 7.8$\times10^{-4}$ and 8.1$\times10^{-4}$, resulting in
the final ratio $\sim 18$ between the hadronic molecular state and
the tetraquark state.

In the present experiment at STAR, open charm mesons are
recostructed from their hadronic decay products
\cite{Tlusty:2012ix}. However, a heavy flavor tracker, which will
make the reconstruction of the secondary vertex of open charm
mesons possible, is scheduled to operate in the near future. Then
we are able to find charmed mesons purely coming from the
$X(3872)$ mesons ( $X\to D^{0}\bar{D}^{*0}$ or $X\to
D^{0}\bar{D}^{0}\pi$ \cite{Beringer:1900zz} ) and measure the
yield of $X(3872)$ mesons produced by the coalescence in heavy ion
collisions. The possibility of the $X(3872)$ meson production from
the $B$ meson decay is very low at RHIC top energy. The estimation
on the time evolution of the $X(3872)$ meson abundance shows that
when the number of $D$ mesons observed through the vertex detector
is cumulated to be about $10^4$, at least a few $X(3872)$ mesons
are expected to be produced if the $X(3872)$ meson is in a
hadronic molecule state. If we need to collect one order of
magnitude larger numbers of $D$ mesons to obtain a trace for a
$X(3872)$ meson, we can conclude that we are finding a $X(3872)$
meson in a tetraquark state. Therefore, a factor of 18 smaller
yield for the $X(3872)$ meson in a tetraquark state is enough to
be used to discriminate the structure of the spin-1 $X(3872)$
meson.

\section*{Acknowledgements}

This work was supported by the Korea National Research Foundation
under Grants No. KRF-2011-0020333 and No. KRF-2011-0030621 and the
Korean Ministry of Education through the BK21 program.

\appendix

\section{Dependence of the strong coupling constants on the $X(3872)$
meson mass}

 In order to understand the origin of the big difference
between $g_{X_1D^*D}$ and $g_{X_2D^*D}$ we investigate the
relation between two strong-coupling constants. We simply consider
the decay rate of the $X(3872)$ meson decaying to $D^0$ and
$\bar{D}^{*0}$,

\begin{figure}[!h]
\begin{center}
\includegraphics[width=0.46\textwidth]{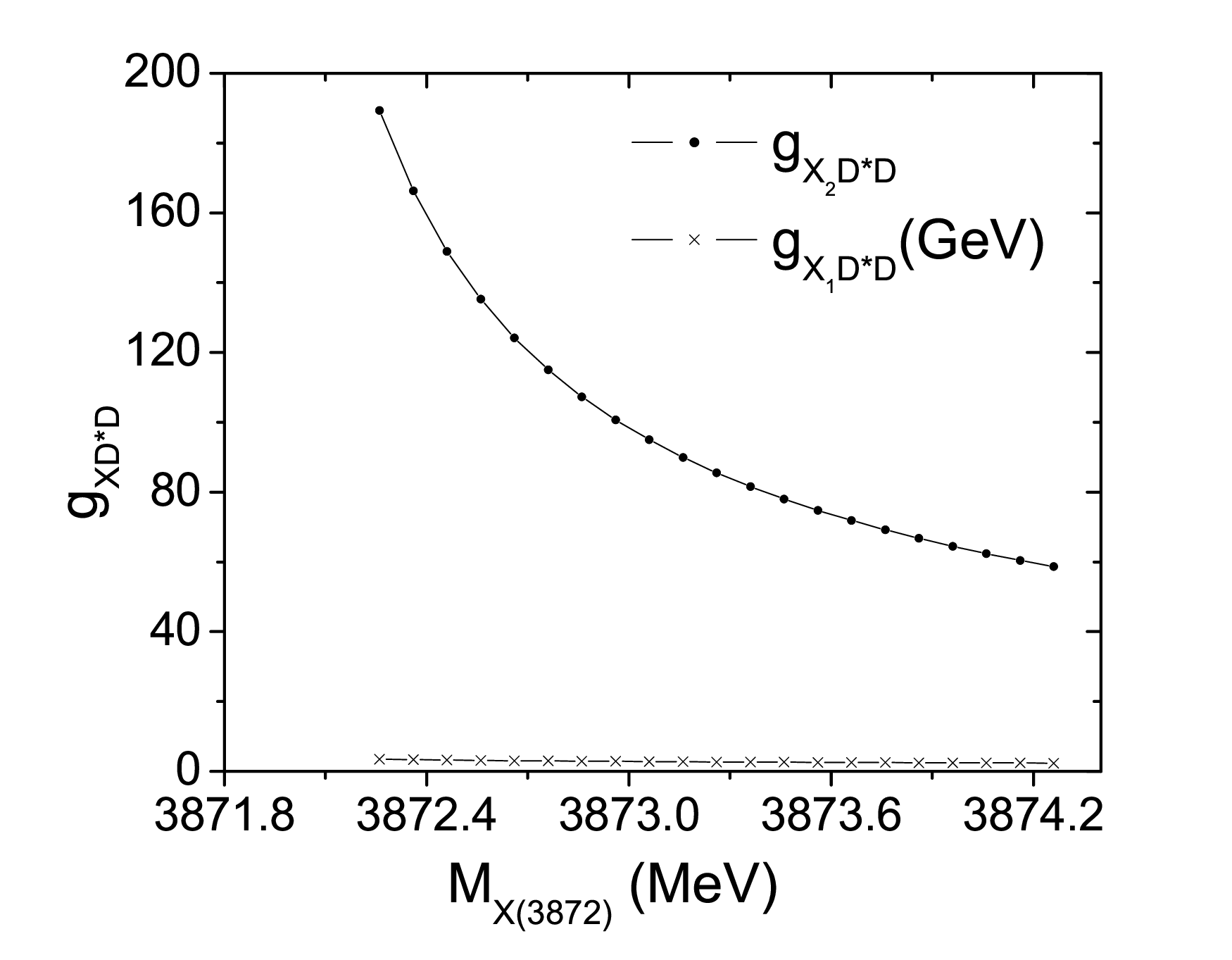}
\end{center}
\caption{The strong coupling constants $g_{X_1D^*D}$ and
$g_{X_2D^*D}$ as functions of the $X(3872)$ meson mass.}
\label{gXD*D}
\end{figure}

\begin{equation}
d\Gamma(X\to D^0\bar{D}^{*0})=\frac{1}{2s_X+1}\frac{(2\pi)^4}
{2m_X}\sum_{pol}|\mathcal{M}|^2d\Phi_2(k;p,q), \label{decay}
\end{equation}
where $d\Phi_2$(k;p,q) is the element of the two-body phase-space
density given by

\begin{equation}
d\Phi_2(k;p,q)=\delta^4(k-p-q)\frac{d^3p}{(2\pi)^32E_{D^0}}\frac{d^3q}
{(2\pi)^32E_{\bar{D}^{*0}}}.
\end{equation}
In the rest frame of the $X(3872)$ meson, the decay rate Eq.
(\ref{decay}) becomes,

\begin{equation}
\Gamma(X\to D^0\bar{D}^{*0})=\frac{1}{8\pi}\frac{1}{m_X^2} |\vec
p|\frac{1}{2s_X+1}\sum_{pol}|\mathcal{M}|^2
\end{equation}
with

\begin{equation}
|\vec p|=\frac{\sqrt{(m_X^2-(m_{D^0}-m_{\bar{D}^{*0}})^2)
(m_X^2-(m_{D^0}+m_{\bar{D}^{*0}})^2)}}{2m_X}.
\end{equation}
Since we consider \textit{two} spin possibilities for the
$X(3872)$ meson to explain \textit{one} experimental observation,
we obtain the following condition using the interaction
Lagrangians, Eq. (\ref{iLagrangians}),

\begin{eqnarray}
& & \quad\frac{1}{3}\sum_{pol}|g_{X_1D^*D}\epsilon^{\mu}(k)
\epsilon_{\mu}^*(q)|^2 \nonumber \\
& & =\frac{1}{5}\sum_{pol}|g_{X_2D^*D}\pi^{\mu\nu}(k)
\epsilon_{\mu}^*(q)p_{\nu}|^2, \label{pspace}
\end{eqnarray}
by requiring $\Gamma_{X_2}=\Gamma_{X_1}$. This condition is
responsible for the difference between $g_{X_1D^*D} = 3.5$ GeV and
$g_{X_2D^*D} = 189$ when a mass of the $X(3872)$ meson is 3872.26
MeV. We can obtain, using Eq. (\ref{pspace}), the other strong
coupling constant of the $X(3872)$ meson when one of them is
known.

We have also found that $g_{X_2D^*D}$ is sensitive to the
variation of the $X(3872)$ meson mass when obtaining it from the
requirement $\Gamma_{X_2}=\Gamma_{X_1}$. Since we know that a
recent measurement of the $X(3872)$ meson mass from its decay
mode, $X\to D^{*0}\bar{D}^0$ is 3872.9 MeV \cite{Adachi:2008sua},
we vary the mass of the $X(3872)$ meson, and see how the strong
coupling constants $g_{X_1D^*D}$ and $g_{X_2D^*D}$ change.

As shown in Fig. \ref{gXD*D}, $g_{X_2D^*D}$ decreases from 189 to
107 whereas $g_{X_1D^*D}$ changes from 3.5 GeV to 2.5 GeV when the
mass of the $X(3872)$ meson increases from 3872.3 MeV to 3872.9
MeV. We expect the above new strong coupling constants obtained
when the mass of the $X(3872)$ meson is 3872.9 MeV to reduce the
cross sections evaluated in Sec. III by a factor of 3 for the
$X_2(3872)$ meson, and by a factor of 2 for the $X_1(3872)$ meson,
respectively.

\section{Hadronic effects on the $X(3872)$ meson by baryons}

In this appendix, we consider a system of baryons interacting with
$X(3872)$ mesons to describe the hadronic effects on the $X(3872)$
meson more realistically. As we find that the cross sections of
the $X(3872)$ meson with different spin states strongly depend on
the strength of the strong coupling-constants, we expect the
hadronic effects on $X(3872)$ mesons by baryons also lead to
similar results as shown in Sec. III.

Since the most abundant baryons available in the system are
nucleons, we take the absorption of $X(3872)$ mesons by nucleons
into consideration: $XN\rightarrow \bar{D}^*\Lambda_C$
and $XN \rightarrow \bar{D}\Lambda_C$. \\

\begin{figure}[!h]
\begin{center}
\subfigure[]{
\includegraphics[width=0.17\textwidth]{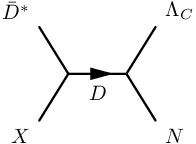}}
\quad \subfigure[]{
\includegraphics[width=0.16\textwidth]{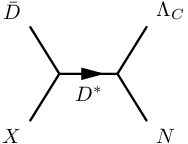}}
\end{center}
\caption{Born diagrams for the $X(3872)$ meson absorption by
nucleons; $XN\rightarrow \bar{D}^*\Lambda_C$ (a) and $XN
\rightarrow \bar{D}\Lambda_C$ (b).} \label{Diagrams2}
\end{figure}
In addition to the interaction Lagrangians introduced in Eq.
(\ref{fLagrangians}), we need the following additional interaction
Lagrangians to describe the diagrams shown in Fig.
\ref{Diagrams2}:

\begin{eqnarray}
{\cal L}_{D^*N\Lambda_C}&=&g_{D^*N\Lambda_C}(\bar{N}\gamma_\mu
\Lambda_C \bar{D}^{*\mu}+D^{*\mu}\bar{\Lambda}_C\gamma_\mu N),
\nonumber \\
{\cal L}_{DN\Lambda_C}&=&ig_{DN\Lambda_C}(\bar{N}
\gamma_5\Lambda_C\bar{D}+D\bar{\Lambda}_C\gamma_5 N),
\label{bLagrangians}
\end{eqnarray}
with coupling constants $g_{D^*N\Lambda_C}=-5.6$ and
$g_{DN\Lambda_C}=13.5$ \cite{Liu:2001ce}. Using these interaction
Langrangians we easily obtain the amplitudes for the reaction,
$XN\rightarrow \bar{D}^*\Lambda_C$,

\begin{eqnarray}
&&{\cal M}_{X_1N\to\bar{D}^*\Lambda_C}=-ig_{DN\Lambda_C}
g_{X_1D^*D}\epsilon^{\mu}_{1}\epsilon_{3\mu}^*\frac{1}{t-m_D^2}
\nonumber \\
&&\qquad\qquad\quad\quad\times\bar{\Lambda}_C(p_4)\gamma_5 N(p_2),
\nonumber \\
&&{\cal M}_{X_2N\to\bar{D}^*\Lambda_C}=-ig_{DN\Lambda_C}
g_{X_2D^*D}\pi^{\mu\alpha}_{1}\epsilon^{*\mu}_{2}(p_1-p_3)_{\alpha}
\nonumber \\
&&\qquad\qquad\quad\quad\times\frac{1}{t-m_D^2}\bar{\Lambda}_C(p_4)
\gamma_5 N(p_2). \label{Matx_XN1}
\end{eqnarray}
for both spin-1 and spin-2 $X(3872)$ meson states. Similarly, we
get the amplitudes,

\begin{figure}[!]
\begin{center}
\includegraphics[width=0.50\textwidth]{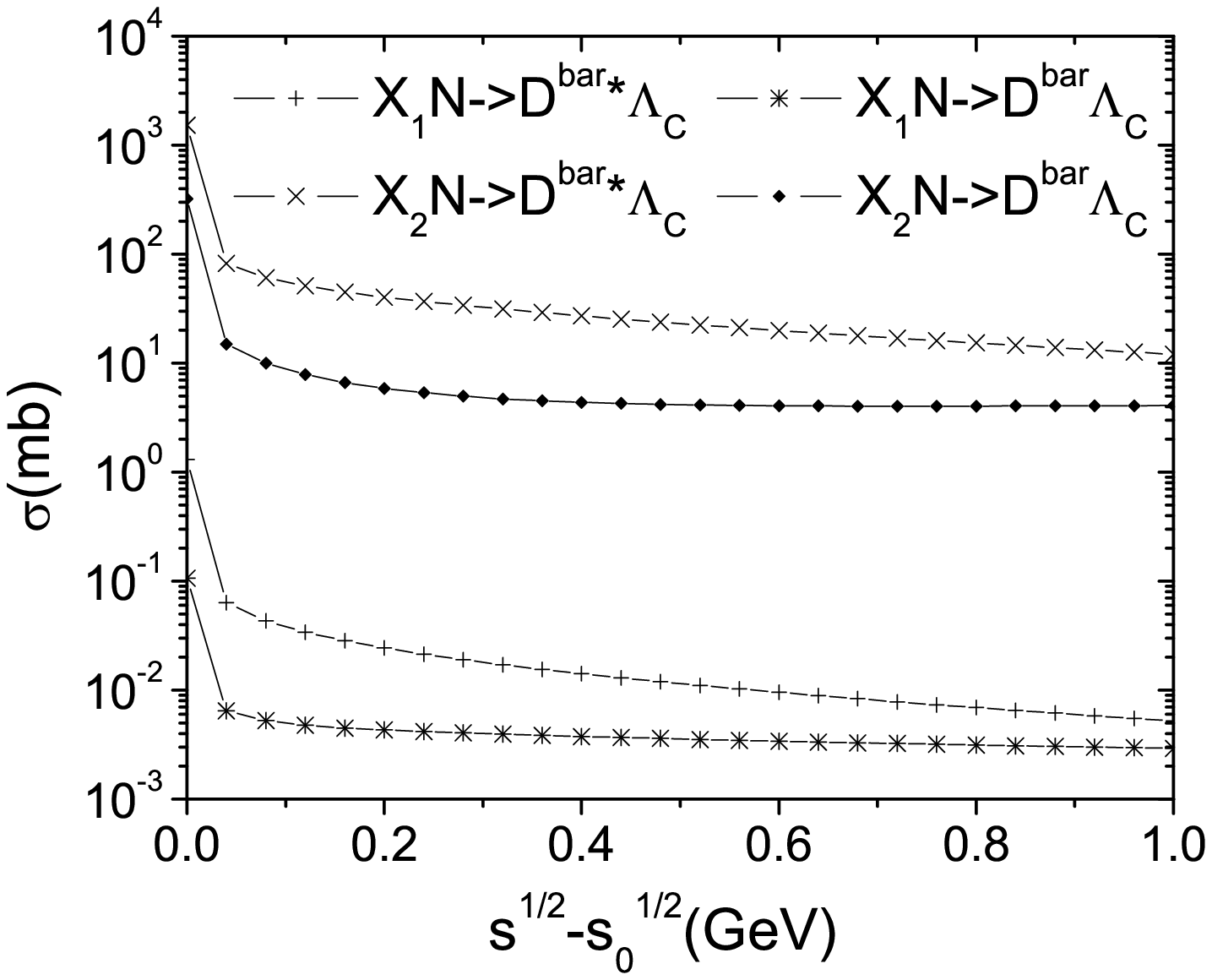}
\end{center}
\caption{Cross sections for the absorption of $X(3872)$ mesons
with different spin states by nucleon via reactions,
XN$\rightarrow \bar{D}^*\Lambda_C$ and XN$\rightarrow
\bar{D}\Lambda_C$. } \label{XNsigma}
\end{figure}

\begin{eqnarray}
&&{\cal M}_{X_1N\to\bar{D}\Lambda_C}=-g_{D^*N\Lambda_C}g_{X_1D^*D}
\epsilon^{\mu}_{1}\frac{1}{t-m_D^{*2}} \nonumber \\
&&\quad\times \Big(-g_{\mu\nu}+\frac{(p_1-p_3)_\mu(p_1-
p_3)_\nu}{m_D^{*2}}\Big)\bar{\Lambda}_C(p_4)\gamma^\nu N(p_2),
\nonumber \\
&&{\cal M}_{X_2N\to\bar{D}\Lambda_C}=-g_{D^*N\Lambda_C}g_{X_2D^*D}
\pi^{\mu\alpha}_{1}p_{3\alpha}\frac{1}{t-m_D^{*2}} \nonumber \\
&&\quad\times\Big(-g_{\mu\nu}+\frac{(p_1-p_3)_\mu(p_1-p_3)_\nu}
{m_D^{*2}}\Big)\bar{\Lambda}_C(p_4)\gamma^\nu N(p_2). \nonumber \\
\label{Matx_XN2}
\end{eqnarray}
for the reaction $XN\rightarrow \bar{D}\Lambda_C$.

We show in Fig. \ref{XNsigma} the final isospin- and spin-averaged
cross sections, (\ref{sigma}) for the above reactions. We see that
cross sections for the spin-1 $X(3872)$ meson are again much
larger than those for the spin-2 $X(3872)$ meson due to the same
reasons discussed in Sec. III. When compared to absorption cross
sections by pions or rho mesons shown in Fig. \ref{X1X2sigma},
absorption cross sections by nucleons for spin-1 $X(3872)$ mesons
are smaller than that of the process, $X_1\pi \rightarrow
\bar{D}^*D^*$, much larger than that for the reaction $X_1\rho
\rightarrow D^*\bar{D}^*$ which contains a three-vector meson
interaction vertex, but similar in size to other cross sections.
For spin-2 $X(3872)$ mesons, absorption cross sections by nucleons
are smaller than those by pions, but much larger than cross
sections by rho meson for reactions, $X_2\rho \rightarrow
D\bar{D}$ and $X_2\rho \rightarrow D^*\bar{D}^*$.

We therefore expect that including hadronic effects on the
$X(3872)$ meson by nucleons accelerates the variation of the
$X(3872)$ meson abundance during the hadronic stage of heavy ion
collisions. However, because the yield of nucleons is smaller than
that of pions by a factor of 10 in the statistical model, the
chance for the $X(3872)$ meson to interact with nucleons is small
and, as a result, hadronic effects on the $X(3872)$ meson by
nucleons would not be dominant. In conclusion, the baryonic
effects on the $X(3872)$ meson is comparable to mesonic effects
but their contribution to the $X(3872)$ meson abundance change in
the hadronic medium would be small due to the smaller yield
compared to that of mesons.

\end{document}